\documentclass[twocolumn,epsfig,superscriptaddress,prb]{revtex4-2}

\usepackage{graphicx}
\usepackage{amssymb}
\usepackage{amsmath}
\usepackage{hyperref}
\usepackage{changes}
\usepackage{comment}
\usepackage{mathtools,nccmath}
\usepackage{ragged2e} 
\usepackage{stackengine}
\usepackage{placeins}
\usepackage[percent]{overpic}
\usepackage{pgfplots}
\usepackage{tabularx}
\usepackage[labelformat=simple]{subcaption}
\usepackage{float}

\DeclareCaptionJustification{justified}{\justifying}
\pgfplotsset{compat=1.8}
\DeclareMathAlphabet{\mathpzc}{OT1}{pzc}{m}{it}

\begin{document}
\title{Signatures of superconducting Higgs mode in irradiated Josephson junctions}

\author{Aritra Lahiri}
\email[]{aritra.lahiri@uni-wuerzburg.de}
\affiliation{Institute for Theoretical Physics and Astrophysics,
University of W\"urzburg, D-97074 W\"urzburg, Germany}

\author{Juan Carlos Cuevas}
\email[]{juancarlos.cuevas@uam.es}
\affiliation{Departamento de F\'{\i}sica Te\'orica de la Materia Condensada, Universidad Aut\'onoma de Madrid, E-28049 Madrid, Spain}
\affiliation{Condensed Matter Physics Center (IFIMAC), Universidad Aut\'onoma de Madrid, E-28049 Madrid, Spain}

\author{Bj\"orn Trauzettel}
\affiliation{Institute for Theoretical Physics and Astrophysics, University of W\"urzburg, D-97074 W\"urzburg, Germany}
\affiliation{W\"urzburg-Dresden Cluster of Excellence ct.qmat, Germany}
\date{\today}

\begin{abstract}
{The Higgs mode, originally proposed in the context of superconductivity, corresponds to oscillations of the amplitude 
of the superconducting order parameter. Recent THz-domain optical studies have found signatures consistent with the 
Higgs mode, but its unambiguous detection is still challenging. We predict that the existence of the Higgs mode 
can be unambiguously revealed by standard measurements of the transport characteristics in microwave-irradiated asymmetric
and transparent Josephson junctions. One signature of the Higgs mode in a Josephson junction is the microwave-induced enhancement 
of the second harmonic of the equilibrium current-phase relation (at zero DC bias voltage), whose sign differs from its 
expected value in the absence of the Higgs mode. As the radiation frequency is varied, this enhancement exhibits resonant 
behavior when the microwave frequency is tuned across the Higgs mass. The second signature that we propose is the 
enhancement of the second harmonic of the AC Josephson current at finite DC voltage bias, which can be probed in a 
customary analysis of the Shapiro steps in a microwave-irradiated junction.}
\end{abstract}
\maketitle

\section{Introduction}
Superconductors (SCs) are characterised by the spontaneous breaking of global $U(1)$ symmetry, resulting in a complex 
order parameter (OP) $\Delta(t) = |\Delta(t)|\exp(i\vartheta(t))$. In equilibrium, the amplitude of the OP, $|\Delta(t)|$, 
is static and results in a spectroscopic gap. Nevertheless, in a non-equilibrium dynamical scenario, it embodies a 
collective mode, namely, the Higgs mode~\cite{Higgs1964,Littlewood1982,Varma2002,Pekker2015,Shimano2020,Matsunaga2012,
Matsunaga2013,Matsunaga2014,Katsumi2018,Beck2013,Matsunaga2017}. Unlike the Nambu-Goldstone mode associated with the 
phase $\phi(t)$, which gets screened by electromagnetic fields and acquires a mass on the order of the plasma 
frequency~\cite{Anderson1958}, the Higgs mode is electrically neutral which makes its detection challenging. Only 
recently, non-linear electromagnetic excitation with THz spectroscopy have observed signatures consistent with the Higgs
mode~\cite{Matsunaga2012,Matsunaga2013,Beck2013,Matsunaga2014,Katsumi2018,Sherman2015,Chu2020,Vaswani2021}. 

Alternatively, there have been a few theoretical studies predicting signatures of the existence of the Higgs mode in 
transport measurements~\cite{Silaev2020,Tang2020,Vallet2023,Plastovets2023,Heckschen2022,Kuhn2024,Vallet2024}, with 
some of them exploring a Josephson setup~\cite{Lee2023,Lahiri2025,Vallet2024}. The Josephson effect~\cite{Josephson1962}, 
characterized by the coherent tunneling of Cooper pairs across a Josephson junction (JJ) of two superconducting leads, 
is a hallmark of phase coherence of the SCs. As such, the Josephson current serves as a proxy for the SC OP, and it may 
be expected to bear an imprint of the SC collective modes~\cite{Pekker2015,Lee2023}. In particular, 
Refs.~\cite{Scalapino1970,Pekker2015} have argued that the Josephson coupling between two SCs in a voltage ($V$) biased 
JJ, which bears the form $\sim \Delta_{0,L}\Delta_{0,R}\cos(\phi)$ with $\Delta_{0,L/R}$ being the equilibrium gaps of 
the left($L$)/right($R$) leads and $d\phi/dt=d/dt(\vartheta_L-\vartheta_R)=2eV$ is 
the Josephson phase (we use $\hbar=1$), provides a way to excite the Higgs mode at frequency equalling $2eV$. Subsequently, 
Ref.~\cite{Lahiri2025} proposed that the resulting signatures of the Higgs mode may be found in the AC Josephson 
current in voltage-biased JJs with high transparency even in the absence of external irradiation. Unlike the usual 
Josephson current, which oscillates at the Josephson frequency $\omega_J=2eV$ for a constant DC voltage 
bias $V$, the Higgs mode manifests as a time-dependent component in the OP, which enhances the second harmonic Josephson 
current at frequency $2\omega_J$. In junctions with highly asymmetric/unequal SC gaps on the two leads, along with 
high transparency, this Higgs-enhanced $2\omega_J$ current may even dominate the $\omega_J$ current, which
constitutes a clear indication of the Higgs mode in conventional JJs with s-wave SCs, in the absence of time-reversal 
symmetry breaking~\cite{trsnote,Sellier2004,Sickinger2012,Goldobin2007,Frolov2006,Li2019,Yokoyama2014,Tanaka1997,
Ng2009,Asano2003,Trimble2021,Can2021,Tummuru2022,Sun2024}.

The problem with these proposals is that a direct detection of such a Higgs-enhanced Josephson current is virtually 
impossible due to the very high required frequency [the Higgs frequency is set by the superconducting gap, which for 
Al SCs as an example would be on the order of $45$~GHZ]. To circumvent this problem we propose to follow the conventional 
way in which the AC Josephson effect is revealed, namely, via the measurement of the transport characteristics under
microwave irradiation \cite{Baronebook,Cuevas2002,Chauvinthesis,Shapiro1963,Kot2020,Siebrecht2023}. To be precise,
we propose to investigate highly asymmetric and fairly transparent Josephson junctions irradiated by microwaves at frequency 
$\omega_r$. We predict the following two unambiguous signatures of the existence of the Higgs mode in a JJ:
First, in the absence of a DC voltage bias and for intermediate transparencies, the Higgs mode induces a significant 
enhancement of the second harmonic of the microwave-assisted current-phase relation (CPR), which is otherwise typically 
dominated by the term $\sim\sin(\phi)$. On tuning $\omega_r$ across $\omega_H$, this Higgs-induced second harmonic 
component displays a resonant behavior, reflecting the underlying Higgs physics. Notably, the Higgs mode induces a sign 
change in the second harmonic of the CPR, which manifests in the full CPR as well. Second, when such asymmetric and transparent irradiated JJs are subjected to a DC voltage bias as well, besides the usual AC Josephson effect where the current develops harmonics of the Josephson 
frequency $\omega_J$ associated with the DC component of the voltage, it also develops harmonics of the radiation 
frequency. For voltages equalling rational multiples of the radiation frequency, $\omega_J=(n/m)\omega_r$ where $n,m$ 
are integers, these components synchronize and constructively add to yield a DC current~\cite{Shapiro1963,Baronebook,Cuevas2002}, 
appearing as Shapiro spikes/steps (SS) in the DC current-voltage characteristics (IVC). Crucially, the amplitudes of 
the SSs carry an imprint of the AC components of the Josephson current in the \emph{absence} of the radiation. In this
work, we extensively discuss these two signatures with the help of a microscopic theory that accounts for the 
self-consistent dynamics of the order parameter in microwave-irradiated junctions. With this theory, we demonstrate 
that microwave-assisted CPR, as well as Shapiro steps in a DC IVC measurement, both provide viable ways to detect the Higgs renormalized $2\omega_J$ Josephson current. Lastly, we remark that while Ref.~\cite{Vallet2024} had considered a seemingly similar setup, it has some key 
differences. In Ref.~\cite{Vallet2024}, the bulk of the superconducting leads constituting the JJ are subjected 
to a uniform oscillating vector potential to induce the Higgs oscillations, which show up in the Josephson current. 
However, in typical junctions, the microwave radiation manifests as an oscillating voltage across the barrier, 
which is not equivalent to an oscillating vector potential permeating the leads. As we demonstrate below, this 
difference gives rise to qualitatively distinct OP dynamics, and hence, different CPR and SSs.

The rest of this manuscript is organized as follows: In Sec.~\ref{phenom}, we start with a phenomenological model to exlain the origin of the Higgs oscillations in microwave-irradiated JJs. We specialize to the case of microwave-assisted CPR in Sec.~\ref{phenomcpr}, followed by an exposition of SSs in Sec.~\ref{phenomss}, demonstrating in both cases how the Higgs mode manifests in the current. After a description of our microscopic Floquet-Keldysh formalism in Sec.~\ref{micromodel} employed to rigorously analyze the Higgs signatures, we present the numerical results in Secs.~\ref{resulcpr} and~\ref{resulss}, exploring in detail the microwave-assisted CPR and SSs, respectively. Finally, we summarize our conclusions, and discuss some practical considerations in Sec.~\ref{discconc}.

\section{Phenomenology}
\label{phenom}
Before discussing the microscopic theory that allows us to quantitative predict how the Higgs mode shows up in the 
transport characteristics of irradiated JJs, we illustrate in this section the underlying principles 
with a phenomenological model. For this purpose, we first recapitulate the key results of Ref.~\cite{Lahiri2025}. 
Conventional treatments of JJs employ BCS mean-field SC Hamiltonians with static/time-independent gap amplitudes. 
In contrast, on relaxing this constraint, we find that under suitable conditions, a self-consistent 
solution for the OP in JJs driven by microwave radiation and/or DC voltage bias acquires a time-dependent \emph{amplitude} 
governed by the Higgs mode. Crucially, this should be distinguished from the more familiar case where the time-dependence 
of the OP arises from a time-dependent \emph{phase} but a static amplitude. Similar to the usual equilibrium/static 
proximity effect wherein the gaps of two SCs in contact with each other is modified in the vicinity of the junction, these 
Higgs oscillations may be understood as a dynamical proximity effect between the two voltage-biased SCs. We demonstrate 
this with a phenomenological model of the two OPs coupled at the junction. We relegate the details to Appendix~\ref{appA}, and 
summarize the key results here. Assuming without loss of generality that the left SC is at a potential $V(t)$ relative to 
the right SC, on decoupling the two SCs constituting the JJ, the OPs have the form $\Delta_{0,L/R}e^{-i\vartheta_{L/R}(t)}$ 
where $\vartheta_{L}(t)=\vartheta_{0,L}+\int_{-\infty}^t 2eV(t')dt'$ and $\vartheta_R(t)=\vartheta_{0,R}$~\cite{Anderson1965,pairoscnote}. 
This relative time-depedence imprints upon the leading order coupling between the OPs of the two SCs $\sim J\Re(\Delta_L(t)^*\Delta_R(t))$ 
where $J\sim\mathcal{T}^2$ with $\mathcal{T}$ being the tunnel coupling. For a DC voltage bias, this provides a stimulus at 
frequency $2eV$~\cite{Scalapino1970}, which leads to the usual AC Josephson effect~\cite{Josephson1962}. Additionally, the 
same time-dependence enhances the usual proximity effect by elliciting a dynamical response; along with a change in the static 
gap amplitude, the OP also acquires a time dependent part whose spectrum derives from the time-dependence of the coupling. 
Throughout this work, we consider highly asymmetric JJs with $\Delta_{0,L}<\Delta_{0,R}$ without loss of generality, where 
the external drive excites the Higgs mode only in the left SC. Expanding the OP amplitude in the left SC as 
$\Delta_{L}(t,x)=\Delta_{0,L}(x)+\delta\Delta_{L}(t,x)$, with $\delta\Delta_{L}(t,x)$ capturing the time dependence, and 
the left lead occupying $x\leq 0$, we find that to the leading order in $J$~\cite{Lahiri2025}
\begin{subequations}
\begin{align}
\delta\Delta_{L}(\omega,q)=&\chi_{L}(\omega,q)X_{L}(\omega),\\
X_L(\omega)=&\int \frac{dt }{2\pi} e^{i\omega t}2J\Delta_{0,R}\cos\big(\phi(t)\big),\\
\chi_L(\omega,q)=&\frac{1}{-(\omega+i0)^2+\omega_{H,L}(q)^2}.
\end{align}\label{deldeltageneral}
\end{subequations}
Here, $\phi(t)=\vartheta_L(t)-\vartheta_R(t)$ is the Josephson phase, $X_L(\omega)$ captures the dynamic excitation provided 
by the coupling between the two OPs, and $\chi_L(\omega,q)$ is the Higgs susceptibility of the left SC which is peaked at 
the Higgs dispersion frequency $\omega=\omega_{H,L}(q)=2\Delta_{0,L}+c^2q^2$ where $q$ is the wavevector and $c$ is the 
Higgs velocity. A detailed treatment of Higgs-mode damping lies beyond the scope of this phenomenological analysis. 
Consequently, the present results are not expected to remain valid when the drive frequency $\omega$ approaches or exceeds 
$\omega_{H,L}$. This limitation, however, does not apply to the microscopic numerical analysis which follows this section. 
Nevertheless, we find that certain key qualitative features are captured by our phenomenological analysis. As introduced 
earlier, we now propose two ways to excite the Higgs mode: (i) phase bias, and (ii) DC voltage bias, with both cases involving a microwave irradiation. These are described in the following subsections. 

\subsection{Phase bias: Current-phase relation}
\label{phenomcpr}
Let us first assume that the JJ is subjected to microwave radiation in the presence of a phase bias, but in the absence of a DC voltage bias. The radiation imposes an AC voltage $V_{AC}\cos(\omega_r t)$~\cite{Baronebook,Cuevas2002,Chauvinthesis,Shapiro1963,Kot2020,Siebrecht2023}, 
which corresponds to the Josephson phase
\begin{align}
\phi(t) = \phi_0 + \underbrace{(2eV_{AC}/\omega_r)}_{\coloneqq \alpha}\sin(\omega_J t),\label{Jphase}
\end{align}  
where $\phi_0$ denotes the phase bias. Henceforth, we use $\alpha=2eV_{AC}/\omega_r$ to parametrize the strength of the 
radiation. Before turning to the Higgs response, we first examine the trivial case where it is absent. We start with the 
``adiabatic" approximation (AA) \cite{Baronebook,Bergeret2010}. The starting point in this approximation is the 
expression for the equilibrium Josephson current in the absence of microwaves is given by $I=\sum_n I^{(n)}\sin(n\phi_0)$. 
A microscopic analysis shows that the harmonics $I^{(n)}$ progressively decrease with increasing $n$, $|I^{(n)}|>|I^{(n+1)}|$, 
and contribute noticeably only at high transparencies. Additionally, their signs alternate, $\text{sgn}(I^{(n)}) = 
-\text{sgn}(I^{(n-1)})$, with $I^{(1)}>0$, $I^{(2)}< 0$, and so on. The leading term, 
$I^{(1)}\sim J\Delta_{0,L}\Delta_{0,R}$, determines the critical current in the tunnel limit. The AA that describes the supercurrent in the presence of microwaves simply consists of substituting the time-dependent phase of Eq.~\eqref{Jphase} in the equilibrium 
CPR. This leads to the generalized result: $I = \sum_n \bar I^{(n)} \sin(n\phi_0)$, where the amplitudes $\bar I^{(n)}$ 
incorporate the effect of the microwaves and are related to the equilibrium amplitudes $I^{(n)}$ as we specify below. 
This type of approximation only works for slowly varying $\phi(t)$, which permits us to neglect the frequency dependence 
of $\bar I^{(n)}$ and replace them with their zero-frequency values \cite{Baronebook,Bergeret2010}. 
In the presence of microwave radiation, we obtain on using Eq.~\eqref{Jphase} that 
\begin{align}
I=& \sum_n \sum_a I^{(n)}J_a(n\alpha)\sin(n\phi_0 + a\omega_r t). \label{IAA1}
\end{align}
The DC supercurrent follows as
\begin{align}
I=& \sum_n  \underbrace{I^{(n)}J_0(n\alpha)}_{\bar{I}^{(n)}} \sin(n\phi_0), \label{IDCAA}
\end{align}
defining $\bar{I}^{(n)}=I^{(n)}J_0(n\alpha)$. Specifically, we note that the second harmonic component, $\bar{I}^{(2)}$, depends on $\alpha$ as $J_0(2\alpha)$, and $\bar{I}^{(2)}(\alpha=0)<0$.

Returning to the case with Higgs renormalization, we shall now see that the $J_0(2\alpha)$ profile, as well as the sign of $\bar{I}^{(2)}(\alpha=0)$ obtained above for the second harmonic current is altered significantly. Plugging Eq.~\eqref{Jphase} in Eq.~\eqref{deldeltageneral}(a--c), we obtain the change in the OP as
\begin{subequations}
\begin{align}
\delta\Delta_L(t,x)=&\sum_{m} \delta\Delta_{L,m}(x) e^{-im\omega_r t},\\
\delta\Delta_{L,m}(q)=& \begin{cases}
\cos(\phi_0) J_m(\alpha) \chi_L(m\omega_r,x)  &\!\!\! m:\text{even}\\
-i\sin(\phi_0) J_m(\alpha) \chi_L(m\omega_r,x)  &\!\!\! m:\text{odd}
\end{cases},\\
\chi_L(\omega,x)=&\sum_q \chi_L(\omega,q)e^{iqx}=\frac{e^{-\frac{\lvert x\rvert \sqrt{\omega_{H,L}^2 - \omega^2}}{c}}}{2c{\sqrt{\omega_{H,L}^2-(m\omega_r)^2}}} .
\end{align}\label{deltaDeltazerov}
\end{subequations}
This shows that the OP contains harmonics of $\omega_r$, which are resonantly excited when the harmonic frequency 
$m\omega_r$ matches the Higgs frequency $\omega_H$. The higher harmonics decay parametrically due to the factor 
$J_m(\alpha)$. We reiterate that this analysis is only valid for small $\omega_r\ll\omega_{H,L}$, along with $\alpha<1$, 
due to the limitations of the AA and the phenomenological field theory for the coupled OPs. Within the AA, promoting the 
equilibrium supercurrent to $I\sim J\Delta_L(t)\Delta_{0,R}\sin(\phi(t))$ and using Eq.~\eqref{deltaDeltazerov}, the 
DC supercurrent once again bears the form given by Eq.~\eqref{IDCAA}, albeit, with the Higgs-renormalized amplitudes $\bar{I}^{(n),\text{Higgs}}$. In particular, while the first harmonic remains unchanged, the 
amplitude of the second harmonic, which encodes the Higgs renormalization, becomes
\begin{align}
\bar{I}^{(2),\text{Higgs}} \sim &J \sum_m (-1)^m \frac{[J_m(\alpha)]^2}{\sqrt{\omega_{H,L}^2-(m\omega_r)^2}}. \label{Iup2Higgs}
\end{align}
From this, we can distinguish the Higgs resonance when harmonics of $\omega_r$ equal $\omega_{H,L}$, and the dependence on $\alpha$. We find that the latter is equally helpful in indicating the underlying Higgs-renormalization. Importantly, along with the difference in the $\alpha$-dependence with [Eq.~\eqref{Iup2Higgs}] and without [Eq.~\eqref{IDCAA}] the Higgs renormalization, we note that $\bar{I}^{(2),\text{Higgs}}(\alpha=0)>0$, in contrast to the Higgs-free case where $\bar{I}^{(2)}(\alpha=0)<0$, as mentioned earlier. This difference in the sign of the second harmonic component plays a significant role in qualitatively altering the CPR skewing the peak current. This is easy to see on retaining only the first two harmonics, yielding the current $I(\gamma,\phi_0)=\bar{I}^{(1)}\sin(\phi_0)+\gamma\sin(2\phi_0)$ (where $\gamma$ is a weighting factor), which satisfies $I(\gamma,\phi_0)=I(-\gamma,\pi-\phi_0)$. Irrespective of the magnitude of the Higgs-renormalized second harmonic component, the qualitative differences in the CPR, arising from the opposite sign of the second harmonic and its distinct $\alpha-$dependence, serve as a clear fingerprint of Higgs renormalization. 

We emphasize that a reliable estimate of the second harmonic amplitude requires a microscopic analysis, since for any finite $\alpha$ multiple harmonics of $\omega_r$ contribute, with the higher ones extending beyond $\omega_{H,L}$ where Higgs-mode damping becomes relevant. Nonetheless, as shown below, the $\alpha$-dependence predicted here agrees with our microscopic numerical results. 

\subsection{Voltage bias: Shapiro Steps}
\label{phenomss}
Next, we consider the case of a microwave-irradiated JJ when a finite DC voltage biased is applied as well. For the discussion
of this problem, it is convenient to first analyze the behavior of the OP and the Josephson current without microwave radiation, 
and then incorporate the effects of the radiation afterward. For a pure DC voltage bias corresponding to $\phi(t)=\phi_0+2eVt$, 
from a similar analysis as above~\cite{Lahiri2025}, we find that as the voltage increases and $\omega_J\to\omega_{H,L}$, the left 
OP modulation is amplified by the Higgs-resonance
\begin{align}
\delta\Delta_{L}(t,x)=& J \chi_L(\omega,x)\cos(\omega_J t+\phi_0),
\end{align}
where $\chi_L(\omega,x)$ is shown in Eq.~\eqref{deltaDeltazerov} above, while the off-resonant $\delta\Delta_{R}$ 
remains negligible. In this case, using the AA for the current as before, we see that it acquires a component oscillating 
at frequency $2\omega_J$, $I_{2\omega_J}=I_{\omega_J}f_H$, where $I_{\omega_J}=J\Delta_{0,L}\Delta_{0,R}$ is the amplitude 
of the usual $\omega_J$ component obtained from the Ambegaokar--Baratoff relation~\cite{Ambegaokar1963}, and 
$f_H \sim 1/\sqrt{\omega_{H,L}^2-\omega_J^2}$ shows a resonant behavior as $\omega_J$ approaches the Higgs frequency. As we show 
below, this Higgs renormalized $I_{2\omega_J}$ may be inferred from Shapiro steps in irradiated JJs.

In the presence of both DC voltage bias and microwave radiation, the net voltage across the junction is given 
by $V+V_{AC}\cos(\omega_r t)$. The corresponding Josephson phase becomes
\begin{align}
\phi(t) = \phi_0 + \omega_J t + \alpha\sin(\omega_J t)\label{Jphase2}
\end{align}
First we consider the case without any Higgs renormalization. Using the AA described above, the current reads
\begin{align}
I(t)=& \; J\Delta_{0,L}\Delta_{0,R}\sin(\phi(t))\nonumber\\
=&\sum_n \underbrace{J\Delta_{0,L}\Delta_{0,R}}_{I_{\omega_J}} J_b(\alpha)\sin(\phi_0+\omega_J t+b\omega_r t).\label{I0}
\end{align}
From this current, we obtain the $b^{\text{th}}$ SS when $\omega_J=|b|\omega_r$, with maximum height 
$SS^1_b=I_{\omega_J}J_{-|b|}(\alpha)$. For convenience, we have defined $SS^a_b$ to refer to the SS at $a\omega_J=b\omega_r$. 
Specifically,
\begin{align}
SS^1_1=2\underset{\phi_0}{\text{max}}\big[I_{\omega_J}J_{-1}(\alpha)\sin(\phi_0)\big]=2I_{\omega_J}J_{-1}(\alpha),\label{SS11}
\end{align}
which has a characteristic $J_{-1}(\alpha)$ profile as a function of $\alpha$. Note that this reveals the amplitude of 
the AC Josephson current at \emph{zero} AC voltage, $I_{\omega_J}$. For high transparencies, equivalently large $J$, 
the higher harmonics $I_{a\omega_J}$ oscillating at frequency $a\omega_J$ may also be significant. Specifically, on 
including the $2\omega_J$ current
\begin{align}
I(t)=&\sum_n I_{\omega_J}J_b(\alpha)\sin(\phi_0+\omega_J t+b\omega_r t)\nonumber\\
+&\sum_{b'} I_{2\omega_J}J_{b'}(2\alpha)\sin(2\phi_0+2\omega_J t+b'\omega_r t),\label{Iac}
\end{align}
two changes occur: (i) The previously obtained $SS^1_b$ are modified as the second line in Eq.~\eqref{Iac} contributes 
to $SS^1_b$ when $2\omega_J=(b'=2b)\omega_r$. This yields
\begin{align}
SS^1_1&=2\underset{\phi_0}{\text{max}}\big[I_{\omega_J}J_{-1}(\alpha)\sin(\phi_0) + 
I_{2\omega_J}J_{-2}(2\alpha)\sin(2\phi_0)\big]. \label{SS11n}
\end{align}
Notice that $I_{2\omega_J}$ introduces a $J_{-2}(2\alpha)$ dependence, deviating from the $J_{-1}(\alpha)$ obtained 
earlier in Eq.~\eqref{SS11}. This is noticeable as $J_{-2}(2\alpha)$ oscillates approximately twice as fast as $J_{-1}(\alpha)$.
(ii) A new set of SSs, $SS^2_b$, are obtained when $2\omega_J=b\omega_r$ for odd $b$. Specifically, $SS^2_1$ has the height
\begin{align}
SS^2_1=2\underset{\phi_0}{\text{max}}\big[I_{2\omega_J}J_{-1}(2\alpha)\sin(2\phi_0)\big]=2I_{2\omega_J}J_{-1}(2\alpha).\label{SS21}
\end{align}
Typically, $I_{2\omega_J}\ll I_{\omega_J}$, particularly in highly asymmetric JJs as we shall see below, which results 
in $SS^2_1\ll SS^1_1$ in the absence of Higgs renormalization.

Returning to the case with Higgs renormalization, the underlying Shapiro physics remains unaltered; what changes is the 
amplitude of $I_{2\omega_J}$. Within the simple AA, with $\Delta_{0,L}\ll \Delta_{0,R}$ and considering a DC voltage resonant 
with the Higgs mode of the left lead $\omega_J\approx 2\Delta_{0,L}$, we obtain $\delta\Delta_L(t)=\delta\Delta_L\cos(\phi(t))$ 
and $\delta\Delta_R\approx0$. The resulting current is given by $I\sim J\Delta_L(t)\Delta_{0,R}\sin(\phi(t))$~\cite{Lahiri2025}. 
Since by arguments of gauge invariance, the left OP responds as $\delta\Delta_L(t)=\delta\Delta_L\cos(\phi(t))$, the Higgs 
renormalized current becomes
\begin{align}
I(t)=&J\big[\Delta_{0,L}+\delta\Delta_L\cos(\phi(t))\big]\Delta_{0,R} \sin(\phi(t))\nonumber\\
=&\sum_b I_{\omega_J}J_b(\alpha)\sin(\phi_0+\omega_J t+b\omega_r t)\nonumber\\
+&\sum_{b'} \underbrace{J\delta\Delta_L\Delta_{0,R}}_{I_{2\omega_J}^{\text{Higgs}}}J_{b'}(2\alpha)
\sin(2\phi_0+2\omega_J t+b'\omega_r t).\label{Iach}
\end{align}
Consequently, on enforcing the condition for $SS^1_1$ along with that of the Higgs resonance, i.e. $\omega_r=\omega_J=2\Delta_{0,L}$, 
we obtain the same expression as Eq.~\eqref{SS11n}, albeit.,with a much stronger Higgs enhancement $I_{2\omega_J}^{\text{Higgs}} \gg 
I_{2\omega_J}$,
\begin{align}
SS^1_1=2\underset{\phi_0}{\text{max}}\big[I_{\omega_J}J_{-1}(\alpha)\sin(\phi_0)+I_{2\omega_J}^{\text{Higgs}}
J_{-2}(2\alpha)\sin(2\phi_0)\big].\label{HiggsSS11}
\end{align}
Similarly
\begin{align}
SS^2_1=2\underset{\phi_0}{\text{max}}\big[I_{2\omega_J}^{\text{Higgs}}J_{-1}(2\alpha)\sin(2\phi_0)\big].\label{HiggsSS21}
\end{align}
Note that $I_{2\omega_J}^{\text{Higgs}}$ depends on the DC voltage. It is peaked at the Higgs resonance $\omega_J \approx 
\omega_{H,L}$. With an experiment probing the SS heights over a range of $V_{AC}$, there are two distinct markers indicative 
of a Higgs enhanced $2\omega_J$ current: (i) The relative magnitudes of $SS^2_1$ and $SS^1_1$ provide an estimate of the strength 
of the Higgs renormalized $I_{2\omega_J}$, which is expected to enhance $SS^2_1$. In conventional JJs without any Higgs 
renormalization, the only other source of $SS^2_1$ is the higher order Josephson effect arising from a large transparency. 
In this work, even though we consider transparencies up to $\approx0.48$, we find that the higher order Josephson effect is 
much weaker than the corresponding Higgs-induced result, particularly in case of highly asymmetric JJs
which, as we show below, suppresses $I_{2\omega_J}$. (ii) A resonantly peaking behavior of $SS^2_1$ on sweeping $\omega_J$ 
across $\omega_{H,L}$ provides an unambiguous signature of the resonant Higgs renormalization of $I_{2\omega_J}$. 
Equivalently, the deviation of $SS^1_1$ from the $J_{-1}(\alpha)$ profile due to the contribution of $I_{2\omega_J}$, which 
introduces a $J_{-2}(2\alpha)$ component, is also expected to show a resonant enhancement across the Higgs resonance.

Let us emphasize that although this simplified analysis captures the essential physics, its validity is restricted to 
tunnel JJs subjected to radiation of sufficiently low frequency and intensity. Noting that the OP varies at the same time 
scale as $\phi$, along with the fact that we consider high transparencies and radiation strength $V_{AC}$, it is imperative 
to account for the proper retarded dynamics when obtaining the microscopic current~\cite{Larkin1967,Ambegaokar1982,
Werthamer1966,Lahiri2023,Cuevas2002,Bergeret2010}. We pursue a microscopic Keldysh formulation in the rest of our work, following 
Refs.~\cite{Lahiri2025} and~\cite{Cuevas2002}. Nevertheless, as we show later, the AA introduced above captures the salient 
features concerning the dependence of the step heights on the radiation strength.

\begin{figure}\includegraphics[width=\columnwidth]{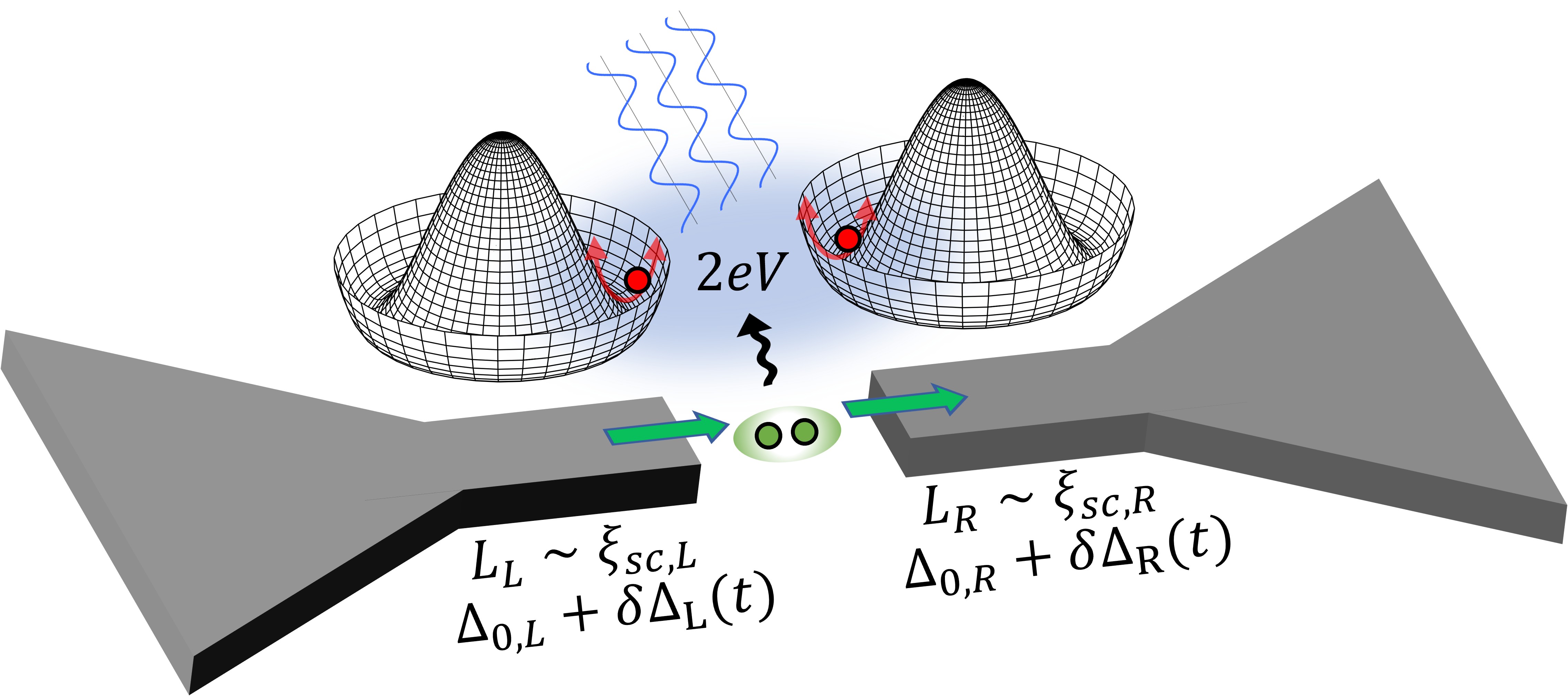}
\caption{Illustration of the JJ, with two leads of length $L_{L/R}\sim\xi_{sc,L/R}$ comparable to the superconducting 
coherence length (subscript $L/R$ denotes left/right) forming a bridge. The outer ends of the leads are connected to 
macroscopic superconducting reservoirs (widening triangles). The JJ has high transparency, and it is highly asymmetric 
with unequal equilibrium gaps, $\Delta_{0,L}\ll\Delta_{0,R}$ without loss of generality. We find that the OP develops a 
time-dependent component, denoted $\delta\Delta_{L/R}(t)$, representing the Higgs mode. This Higgs mode, corresponding to 
radial oscillations of the OP (red balls) in the free-energy landscape of the OP (mexican hat), is excited by radiating 
tunneling Cooper pairs. An external radiation (blue waves) is used to create SSs, which reveal the presence and the 
strength of the Higgs-induced Josephson current.}
\label{Fig1} 
\end{figure}

\section{Microscopic Model} 
\label{micromodel}
Our technical formulation follows from Refs.~\cite{Lahiri2025} and~\cite{Cuevas2002}, based on a self-consistent 
Keldysh-Gorkov framework~\cite{Collado2019,Kamenev2011,Kemoklidze1966,Volkov1974,Vadimov2019,Kuhn2024,Dzero2024}. We, 
nevertheless, describe the crucial aspects to keep the discussion self-contained. We model our junctions with 
s-wave BCS SC Hamiltonians~\cite{Lahiri2025,AGD1975,deGennes1966,Stefanucci2010,Larkin1967,Werthamer1966,
Ambegaokar1982,Lahiri2023,Yeyati1995}. The device, as illustrated in Fig.~\ref{Fig1}, is a superconducting bridge 
comprising two SC leads of length $L$ with $\lambda_F \ll L \sim \xi_{\text{sc}}$, where $\lambda_F$ is the Fermi 
wavelength, and $\xi_{\text{sc}}$ is the SC coherence length. This is because the Higgs oscillations take effect 
over a length scale $\sim \xi_{\text{sc}}$. The leads terminate in macroscopic SC reservoirs~\cite{Samanta1998}. 
For numerical tractability, we consider a single-channel JJ. This is a valid assumption for specular tunneling across 
the interface, which establishes a homogeneous OP along the direction(s) transverse to current transport. This was 
already established in Ref.~\cite{Lahiri2025}, albeit, in the absence of external radiation. The \emph{time-dependent} mean-field Hamiltonian, written in a local 
real-space basis, is given by $H=H_L+H_R+H_{\mathcal{T}}$, where
\begin{align}
H_{L/R}=\sum_{j\in L/R,\sigma} &\big(-\zeta c_{j+1\sigma}^\dagger c_{j\sigma}-\zeta c_{j\sigma}^\dagger 
c_{j+1\sigma}\big) \nonumber \\
+ & \big(\Delta_j(t)c^\dagger_{j\sigma}c^\dagger_{j\sigma'} + \Delta_j^*(t)c_{j\sigma'}c_{j\sigma}\big),
\end{align}\label{Ham}
and
\begin{equation}
H_{\mathcal{T}}=\sum_\sigma \big(W(t)c_{R1\sigma}^\dagger c_{LN_{L}\sigma} + h.c.\big),\ 
W(t)=-\mathcal{T}e^{i\frac{\phi(t)}{2}}.
\end{equation}
Here, $j$ labels sites belonging to the $L$(left) and $R$(right) leads, containing $N_{L}$ and $N_{R}$ sites, respectively. 
Furthermore, $\zeta$ denotes the hopping amplitude, with the bandwidth equalling $4\zeta$, $g>0$ is the BCS attractive 
interaction, $\mathcal{T}$ denotes the junction tunnel coupling, and lastly $\phi(t)$ is given by either Eq.~\eqref{Jphase} or~\eqref{Jphase2} depending on the biasing condition. We remark that by allowing the gaps $\Delta(t)$ to be time-dependent, 
this formulation is able to capture the mean-field dynamics of the OP deriving from the Higgs mode. As described in Appendix.~\ref{appA} following Refs.~\cite{Pekker2015,Lahiri2025}, the Josephson coupling stabilizes the Higgs fluctuations, shifting the OP’s vacuum expectation value from the equilibrium static gap $\Delta_{0}$ to a time-dependent one $\Delta_{0}+\delta\Delta(t)$. The latter is effectively captured by such a time-dependent mean-field formulation~\cite{Volkov1974,Vadimov2019,Kuhn2024,Dzero2024}. Additionally, in the Hamiltonian mentioned above, we have adopted a gauge in which the voltage is shifted from the chemical potentials, and it appears exclusively in the tunneling term~\cite{Cuevas1996}. There are two aspects to the validity of this procedure. First, even in equilibrium, a supercurrent flow establishes a phase gradient within the leads. In our setup, it can be shown that variation of the SC phase in the leads over a distance $\xi_{\text{sc}}$ satisfies $\xi_{\text{sc}}\nabla\phi\sim\tau$, 
where $\tau$ is the normal state transparency. Hence, for highly asymmetric junctions where the requirement of very high 
transparencies may be relaxed, it is justified to neglect the phase variations within the leads~\cite{Lahiri2025}. The 
macroscopic superconducting reservoirs, by virtue of the macroscopic number of transport channels, is immune to this issue. 
Second, beyond equilibrium, the presence of electric fields within the leads could lead to a time-varying phase. We consider 
type I SCs with $\xi_{sc}\gg\Lambda$ where $\Lambda$ is the penetration depth. Since the Higgs and transport physics occur 
over length scales $\sim\xi_{sc}$, it is safe to neglect the fields and the associated phase variations. Hence, 
we assume that the voltage drop is confined to the barrier.

For the DC voltage-biased and irradiated junction, the tunnel amplitude equals
\begin{align}
W(t)=&\sum_{m,n}\underbrace{-\mathcal{T}e^{-i\frac{\phi_0}{2}}\delta_{m,1}J_n\Big(\mfrac{\alpha}{2}\Big)}_{W_{mn}}
e^{-im\frac{\omega_J}{2}t-in\omega_r t},
\end{align}
revealing that energy is transferred in units of $\omega_J/2$ and $\omega_r$. The tunneling self-energy is given by
\begin{subequations}
\begin{align}
\Sigma^{r/a}_{\mathcal{T},RL}(t)&={\Sigma^{r/a}_{\mathcal{T},LR}}^*(t)=W(t)\tau_+ - W^*(t)\tau_-,\\
\Sigma^{r/a}_{\mathcal{T},RL;pq,mn}&=W_{(p-m)(q-n)}\tau_+ - W_{(-(p-m))(-(q-n))}^*\tau_-,
\end{align} 
\end{subequations}
where $\tau$ denotes the Pauli matrices in Nambu space with $\tau_{\pm}=(\tau_0\pm\tau_3)/2$. Since the OP responds to 
the energy supplied by the tunneling pairs, we have
\begin{align}
\delta\Delta_j(t)=&\sum_{m,n}\delta\Delta_{mn;j}e^{-im\frac{\omega_J}{2}t-in\omega_r t},
\end{align}
with the corresponding self-energy being
\begin{subequations}
\begin{align}
\Sigma^{r/a}_{\delta\Delta,j}(t)&=\delta\Delta_j(t)\tau_1,\\
\Sigma^{r/a}_{\delta\Delta;pq,mn;j}&=\delta\Delta_{(p-m)(q-n);j}\tau_1,
\end{align}
\end{subequations}
where $j$ denotes the spatial indices. Therefore, the two-time retarded and advanced Green's functions may be written as 
(see Appendix~\ref{appB})
\begin{align}
G^{r/a}(t,t')=\sum_{p,q} {\int_{-\infty}^\infty} \frac{d\omega}{2\pi} &\overbrace{G^{r/a}
\Big(\omega+p\mfrac{\omega_J}{2}+q\omega_r,\omega\Big)}^{G^{r/a}_{pq}(\omega)} \nonumber \\ 
&\hspace{0.7mm}e^{-i\left(\omega+p\frac{\omega_J}{2}+q\omega_r\right) t+i \omega t'}.\label{Grep}
\end{align}
In this representation, the Dyson equations for the retarded and advanced components become
\begin{align}
G^{r/a}_{pq}(\omega)=&g^{r/a}_{00}(\omega)\delta_{p,0}\delta_{q,0}+\sum_{mn}g^{r/a}_{pq}(\omega)\Sigma^{r/a}_{pq,mn}
G^{r/a}_{mn}(\omega),\label{RADyson}
\end{align}
where all quantities are matrices in Nambu space, and the bare Green's function $g_{pq}(\omega)=g(\omega+p\omega_J/2+q\omega_r)$ 
is defined in the absence of tunneling. The self-energy contains several contributions. Note that $\Sigma_{\mathcal{T}}^< = 
\Sigma_{\delta\Delta}^<=0$~\cite{Cuevas1996}. We also have the reservoir self-energy, $\Sigma^{r/a/<}_{\text{res.};pq,mn}=\zeta^2\tau_3 g_b^{r/a/<} \tau_3 \delta_{p,m}\delta_{q,n}$, where $g_b$ is the boundary Green's function~\cite{Samanta1998}, acting only on the lead sites 
immediately neighboring the reservoir. Lastly, we include the broadening self-energy $\Sigma^{r/a}_{\Gamma;pq,mn}=\mp i(\Gamma/2) 
\delta_{p,m}\delta_{q,n}$ and $\Sigma^{<}_{\Gamma;pq,mn}=i\Gamma f(\omega) \delta_{p,m}\delta_{q,n}$, where $f(\omega)$ is the 
Fermi function. It aids numerical convergence, and accounts for the lifetime arising from, e.g., relaxation to the quasiparticle 
continuum, electron-phonon interaction, etc~\cite{Lahiri2023}. Finally, the lesser Green's function, which is central to our 
calculation, is obtained as~\cite{Jauho1994,Xu2019,Keldysh1964,Stefanuccibook2013,Gonzales2020,Wimmerthesis,Gldecaynote,Collado2019},
\begin{align}
G^<_{mn}(\omega)=& \sum_{r,s} G^{r}_{(r+m)(q+n)}(\omega)\Sigma^<(\omega) {G^{r}}^\dagger_{rs}(\omega),\label{LDyson}
\end{align}
where $G^a_{rs}(\omega)\equiv G^a(w,\omega+r\omega_J/2+s\omega_r)={G^r}^\dagger(\omega+r\omega_J/2+s\omega_r,w) \equiv 
{G^r_{rs}}^\dagger$. 

The non-equilibrium gap equation~\cite{Lahiri2025,Collado2019,Kemoklidze1966,Vadimov2019,Kuhn2024} becomes
\begin{align}
\Delta_j(t)=&iu_{BCS}\big[G^<_{j,j}\big]_{1,2}(t,t)\nonumber\\
\implies \Delta_{mn;j}=&iu_{BCS} \int_{-\infty}^\infty\frac{d\omega}{2\pi}\big[G^<_{mn,00;j,j}\big]_{1,2}(\omega), \label{gapeq}
\end{align}
where $u_{BCS}$ is the BCS interaction in the pairing channel, $j$ denotes the spatial index, $m$ and $n$ denote the harmonics 
of $\omega_J/2$ and $\omega_r$, respectively, and we consider the anomalous component of $G^<$, as evident from the subscript $1,2$.

We self-consistently solve Eqs.~\eqref{RADyson},~\eqref{LDyson} and~\eqref{gapeq} to obtain the OP. Finally, the current is obtained as~\cite{Lahiri2025,Cuevas2002},
\begin{subequations}
\begin{align}
I(t)=&\sum_{mn}I_{mn}e^{-im\frac{\omega_J}{2}-in\omega_r t}\\
I_{mn}=&\sum_{p,q}e \int_{-\infty}^{\infty} \frac{d\omega}{2\pi} \mathbf{tr}\big[ \tau_3\Sigma_{\mathcal{T},LR,(p+m),(q+n)} 
G_{R1LN_L,pq}^{<}(\omega)\nonumber\\
&\hspace{2.5cm}-(L\leftrightarrow R)\big].\label{If}
\end{align}
\end{subequations}
We note that the DC current is obtained as $I_{00}$, while the AC components oscillating at frequency $\omega_J$ and $2\omega_J$ 
are $I_{20}=I_{\omega_J}$, and $I_{40}=I_{2\omega_J}$, respectively. We use these notations interchangeably in this work.

As introduced above, we consider two observables, namely, the CPR, and the SSs. The calculation of the SS amplitudes requires a full account of the Floquet indices corresponding to both DC and AC components of the voltage, exactly following the theory in this section. Recalling that $SS^a_b$ is obtained when $a\omega_J=b\omega_r$, we obtain
\begin{align}
SS^a_b=2\underset{\phi_0}{\text{max}}\Big[ \sum_k & \Im(I_{2ak,-bk}-I_{-2ak,bk})\sin(k\phi_0)\nonumber\\
 + &\Re(I_{2ak,-bk}+I_{-2ak,bk})\cos(k\phi_0) \Big],\label{SSab}
\end{align}
where the first and the second lines correspond to the sine and cosine Josephson currents~\cite{nocosI}.

The calculation of the CPR is simpler; while the procedure described above remains unchanged, the absence of the DC voltage bias ($\omega_J=0$) lets us drop the corresponding Floquet indices altogether, with the time-dependence of all quantities occuring only in harmonics of $\omega_r$. For clarity, we summarise the details. The OP becomes
\begin{align}
\delta\Delta_j(t)=&\sum_{n}\delta\Delta_{n;j}e^{-in\omega_r t}. \label{deltafloquet}
\end{align}
With the self-energies now given by
\begin{subequations}
\begin{align}
\Sigma^{r/a}_{\mathcal{T},RL;q,n}&=W_{(q-n)}\tau_+ - W_{-(q-n)}^*\tau_-,\\
\Sigma^{r/a}_{\delta\Delta;q,n;j}&=\delta\Delta_{(q-n);j}\tau_1,
\end{align}
\end{subequations}
the Green's functions admit the form
\begin{align}
G^{r/a/<}(t,t')=&\sum_{q} {\int_{-\infty}^\infty} \frac{d\omega}{2\pi} \overbrace{G^{r/a/<}
(\omega+q\omega_r,\omega)}^{G^{r/a/<}_{q}(\omega)}\nonumber\\
&\hspace{1.905cm} e^{-i(\omega+q\omega_r) t+i \omega t'}.
\end{align}\begin{figure*}[t]
\begin{subfigure}[b]{0.4\linewidth}
\caption{}\label{subfig:1cpra}
\includegraphics[width=\columnwidth]{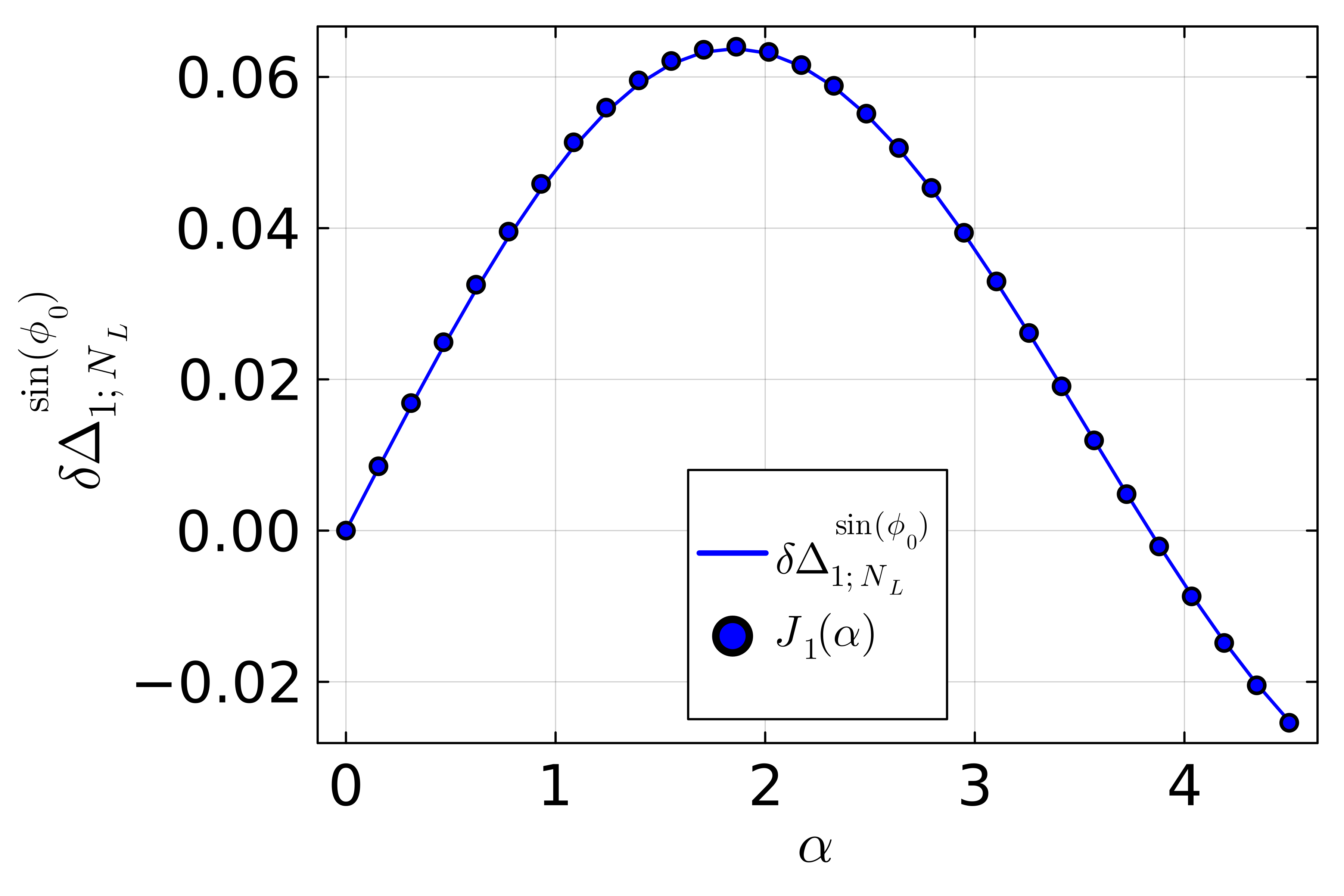}
\end{subfigure}
\begin{subfigure}[b]{0.4\linewidth}
\caption{}\label{subfig:1cprb}
\includegraphics[width=\columnwidth]{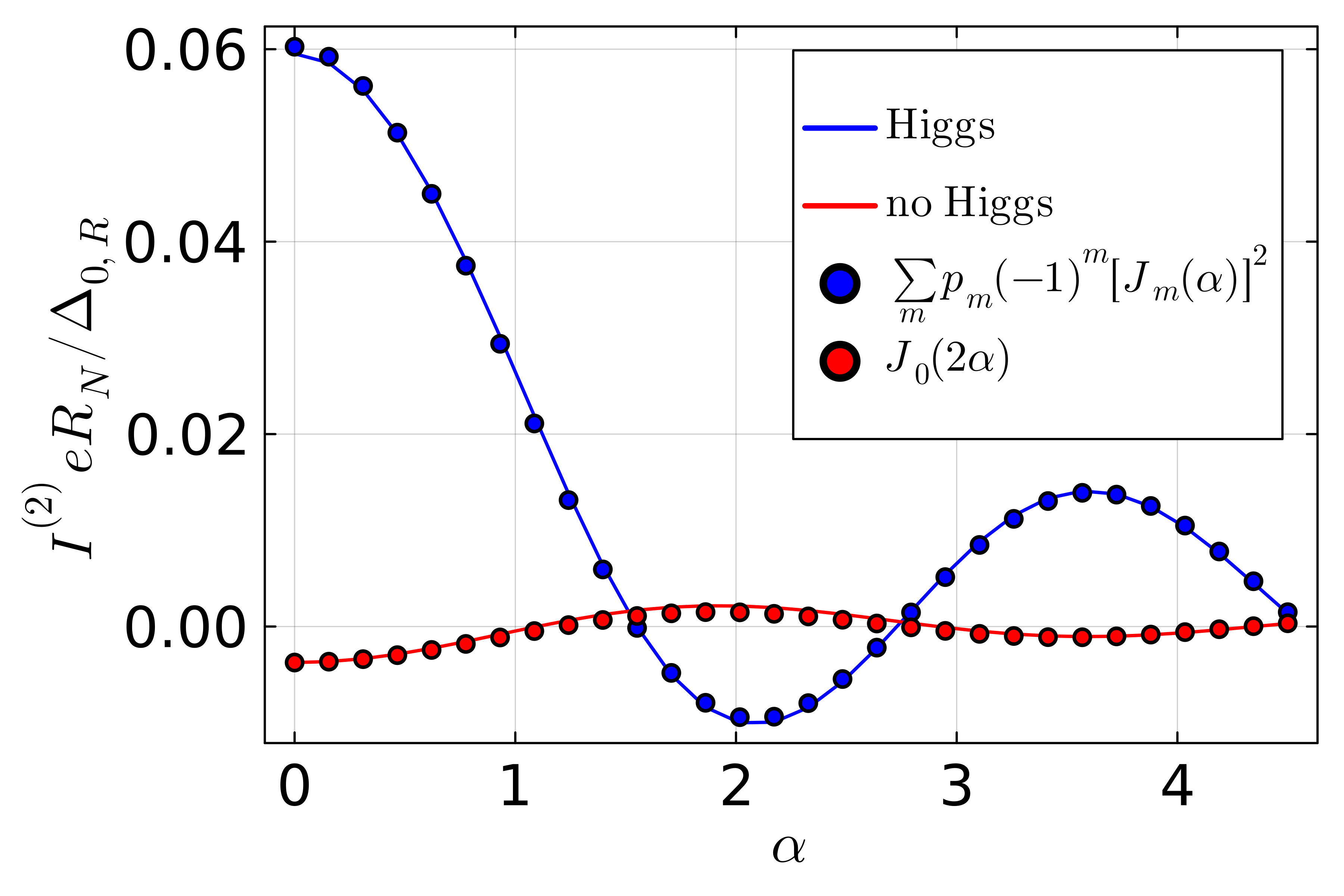}
\end{subfigure}
\begin{subfigure}[b]{0.38\linewidth}
\caption{}\label{subfig:1cprb}
\includegraphics[width=\columnwidth]{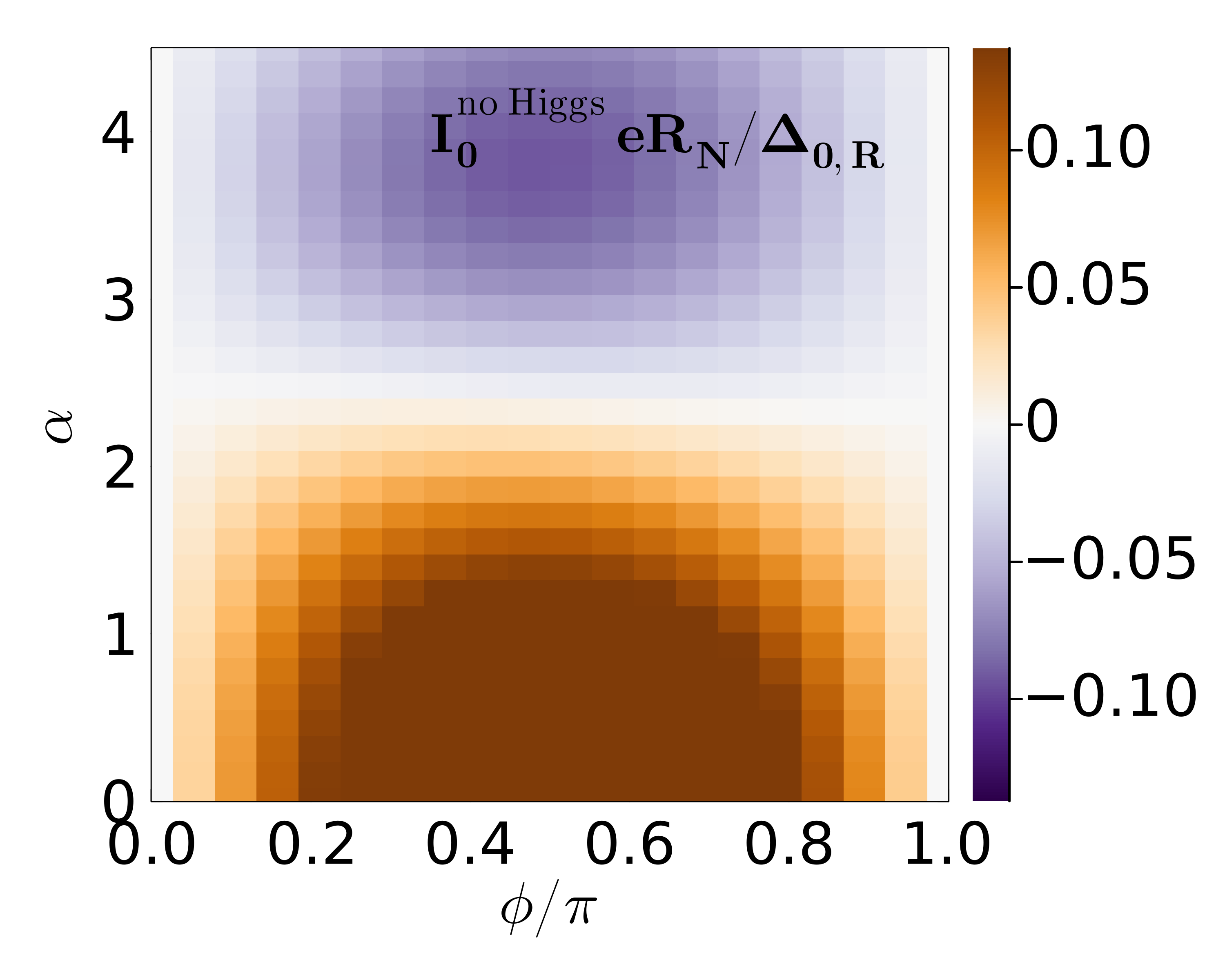}
\end{subfigure}
\begin{subfigure}[b]{0.38\linewidth}
\caption{}\label{subfig:1cprb}
\includegraphics[width=\columnwidth]{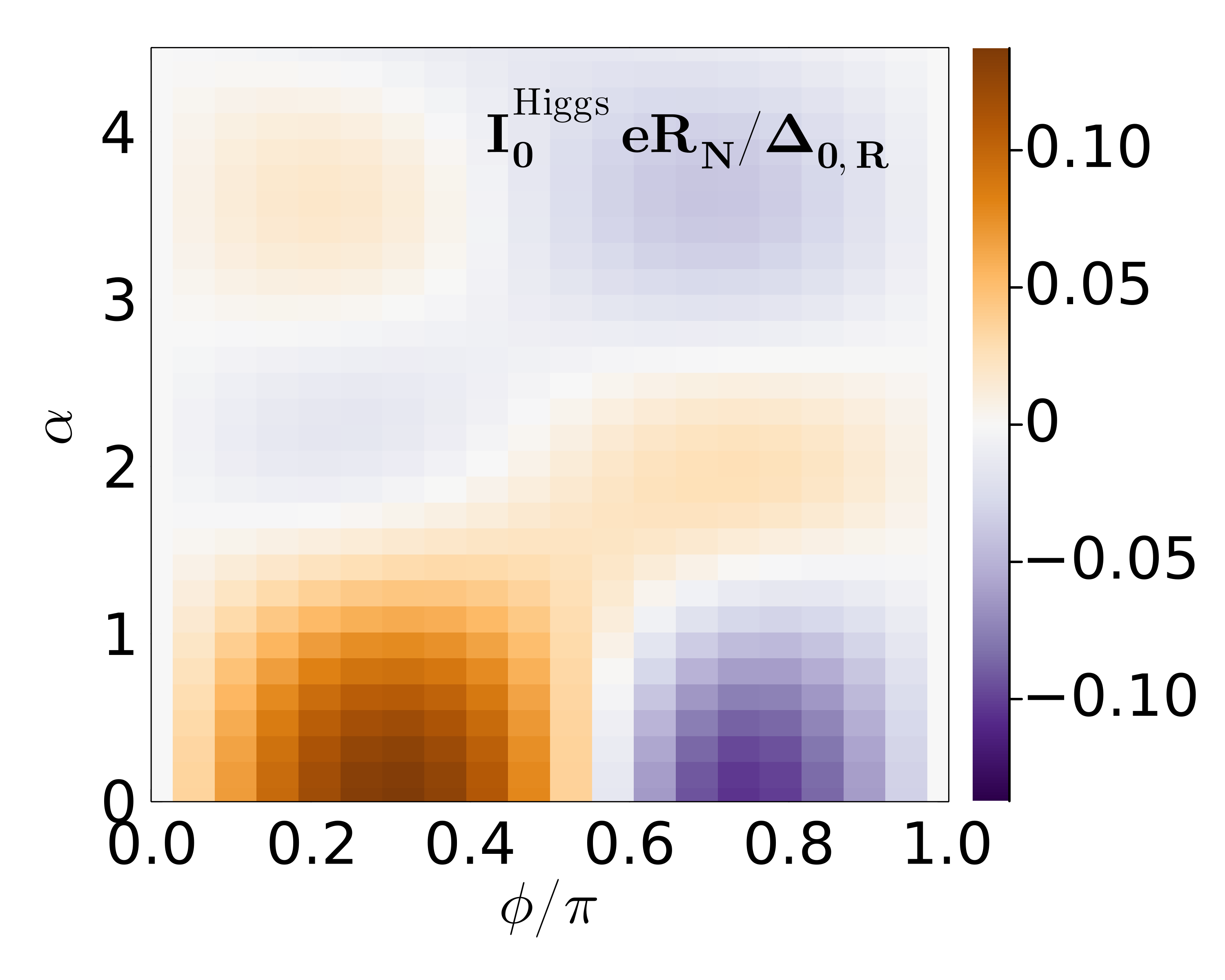}
\end{subfigure}
\caption{(a) Numerically obtained first Floquet harmonic of the OP modulation in the left lead at the junction, $\delta\Delta_{1;j=N_L}$ [$n=1$, $j=N_L$ cf. Eq.~\eqref{deltafloquet}], for $N_L=18$, $N_R=5$~\cite{sizejust}, $\mathcal{T}/\zeta=0.4$ (transparency $\approx0.48$~\cite{Cuevas2002}), $\Gamma=0.0125\Delta_{0,R}=0.0025\zeta$, and $\omega_r\approx 2.29\Delta_{0,L}\approx 1.14\omega_{H,L}$, just over the Higgs resonance. As anticipated from Eq.~\eqref{deltaDeltazerov}, the dominant contribution varies as $\sin(\phi_0)$ (shown here, denoted by the subscript). The $\cos(\phi_0)$ component (not shown) is much weaker. The predicted $J_1(\alpha)$ dependence is also confirmed, as indicated by the blue circles. (b) The second harmonic component of the CPR, $\bar{I}^{(2)}$, as a function of $\alpha$. We present the normalized quantity $\bar{I}^{(2)} e R_N / \Delta_{0,R}$, where $R_N$ is the numerically obtained normal-state resistance. For reference, in symmetric ($\Delta_{0,L}=\Delta_{0,R}=\Delta_{0}$) tunnel JJs $I^{(1)}$ satisfies the Ambegaokar--Baratoff result $I^{(1)} e R_N / \Delta_0 = \pi/2$~\cite{Ambegaokar1963}. In the Higgs-free case (red), obtained by retaining only the zeroth Floquet component of the OP in the self-consistency equation Eq.~\eqref{gapeq}, we confirm the expected $J_0(2\alpha)$ dependence [cf. Eq.~\eqref{IDCAA}]. With Higgs renormalization included (blue), $\bar{I}^{(2),\text{Higgs}}$ is much larger. It starts out positive at $\alpha=0$, in contrast to the Higgs-free case. Moreover, inspired by Eq.~\eqref{Iup2Higgs}, the numerical results are well described by a fit of the form $\sum_m p_m [J_m(\alpha)]^2$, with the coefficients $p_m$ obtained from least-squares regression. We find that $p_{m\geq 5}$ are negligible. (c) We show the CPR $I_0$ with varying $\alpha$, in the absence of Higgs renormalization. Since $\bar{I}^{(2),\text{no Higgs}}$ is negligible, the CPR is not noticeably altered from $I_0\sim I^{(1)}J_0(\alpha)\sin(\phi_0)$. (d) Same as (c), but now we include the Higgs renormalization. Since $\bar{I}^{(2),\text{Higgs}}$ is large, it imparts a $\sin(2\phi_0)$ phase dependence, with the $\alpha-$dependence following from Eq.~\eqref{Iup2Higgs} [see also panel (b)].}
\label{Fig2} 
\end{figure*}Note that, as shown in Appendix~\ref{appB}, this is equivalent to the usual Floquet expansion. Consequently, the current is obtained as
\begin{subequations}
\begin{align}
I(t)=&\sum_{n}I_{n}e^{-in\omega_r t},\\
I_{n}=&\sum_{q}e \int_{-\infty}^{\infty} \frac{d\omega}{2\pi} \mathbf{tr}\big[ \tau_3\Sigma_{\mathcal{T},LR,(q+n)} 
G_{R1LN_L,q}^{<}(\omega)\nonumber\\
&\hspace{2.5cm}-(L\leftrightarrow R)\big].\label{Ifcpr}
\end{align}
\end{subequations}
The CPR is the DC component of the current, obtained from its zeroth Floquet component, $I_0$.

\begin{figure*}[!htb]
\begin{subfigure}[b]{0.38\linewidth}
\caption{}\label{subfig:1cpra}
\includegraphics[width=\columnwidth]{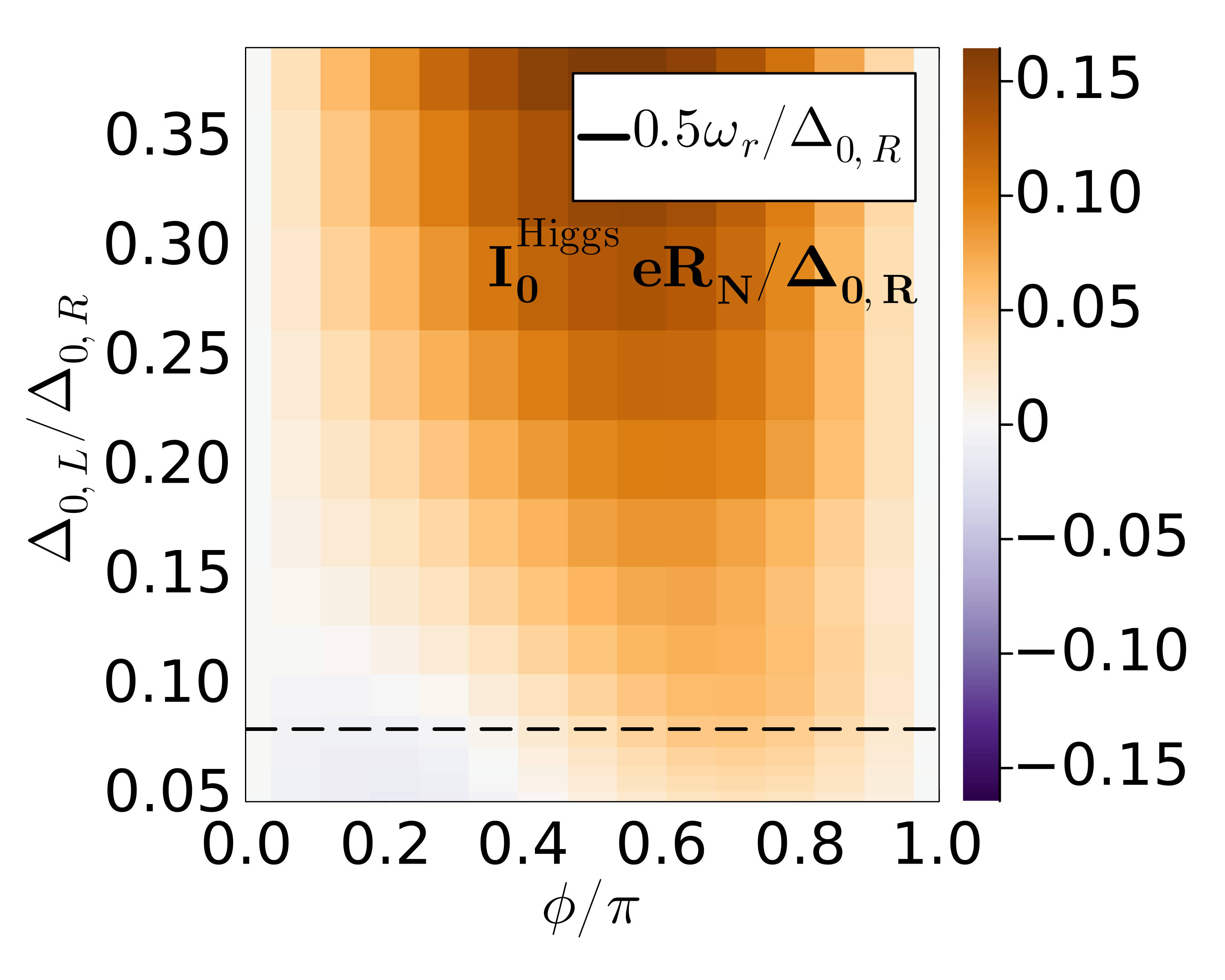}
\end{subfigure}
\begin{subfigure}[b]{0.38\linewidth}
\caption{}\label{subfig:1cprc}
\includegraphics[width=\columnwidth]{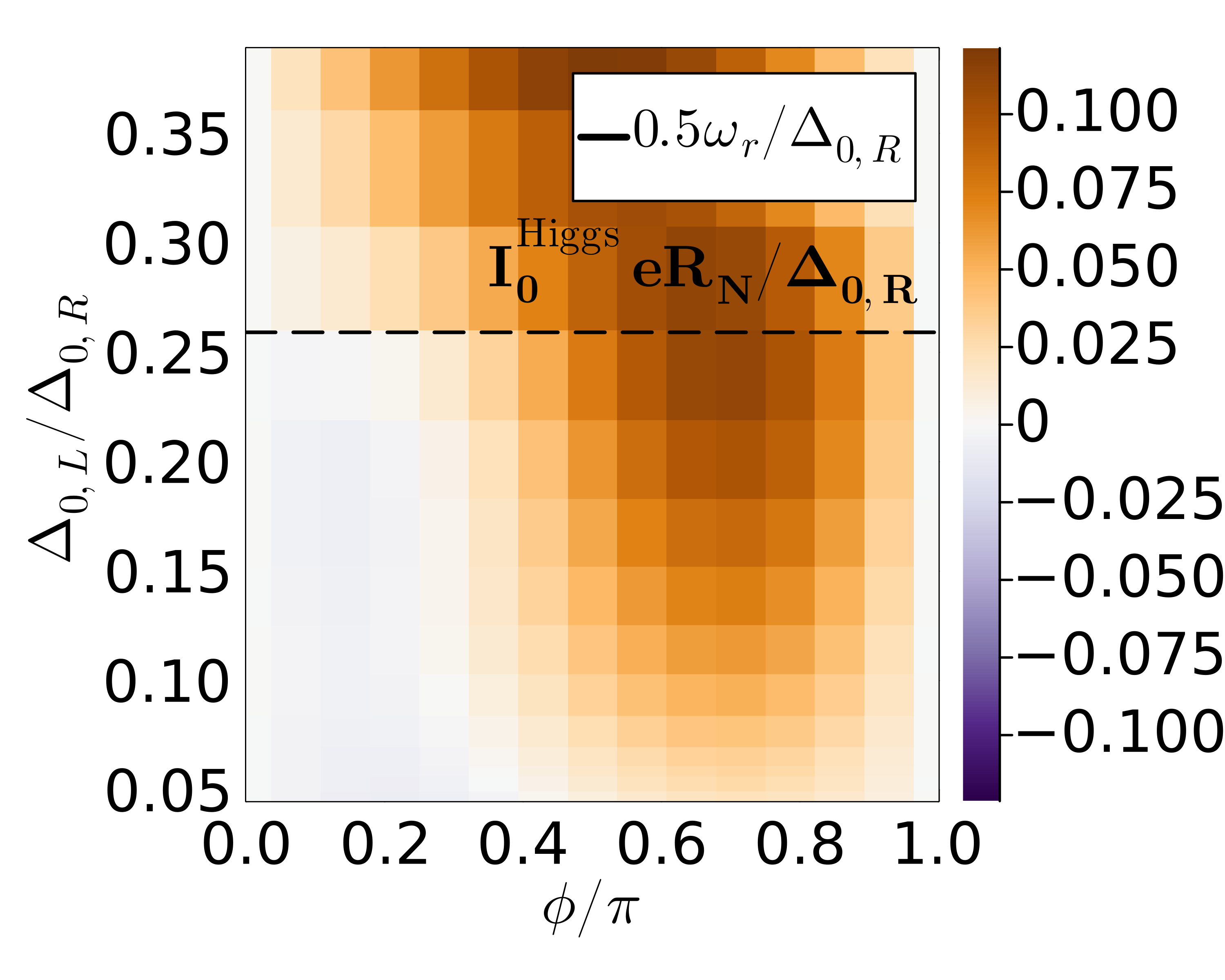}
\end{subfigure}
\begin{subfigure}[b]{0.38\linewidth}
\caption{}\label{subfig:1cpra}
\includegraphics[width=\columnwidth]{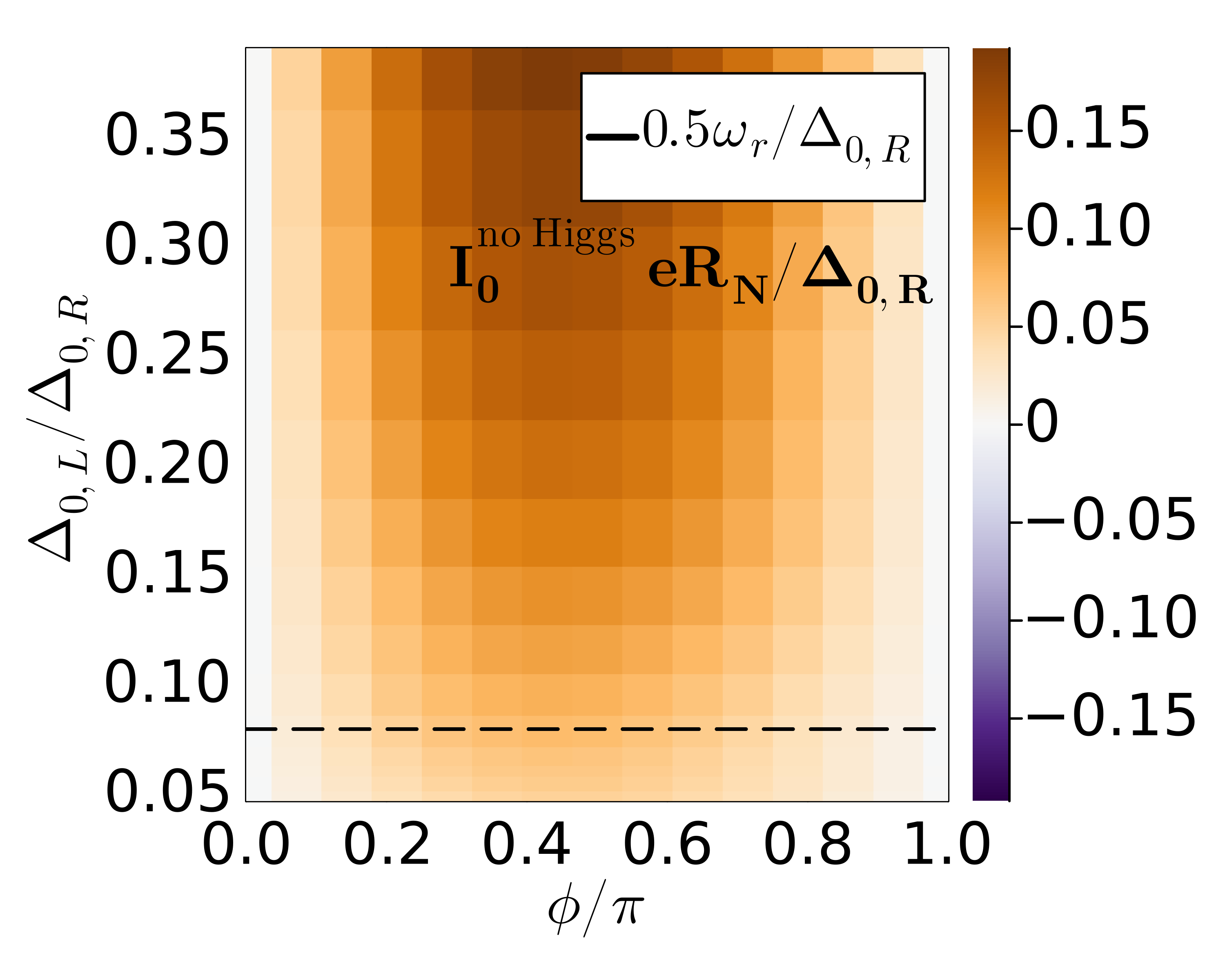}
\end{subfigure}
\begin{subfigure}[b]{0.38\linewidth}
\caption{}\label{subfig:1cprc}
\includegraphics[width=\columnwidth]{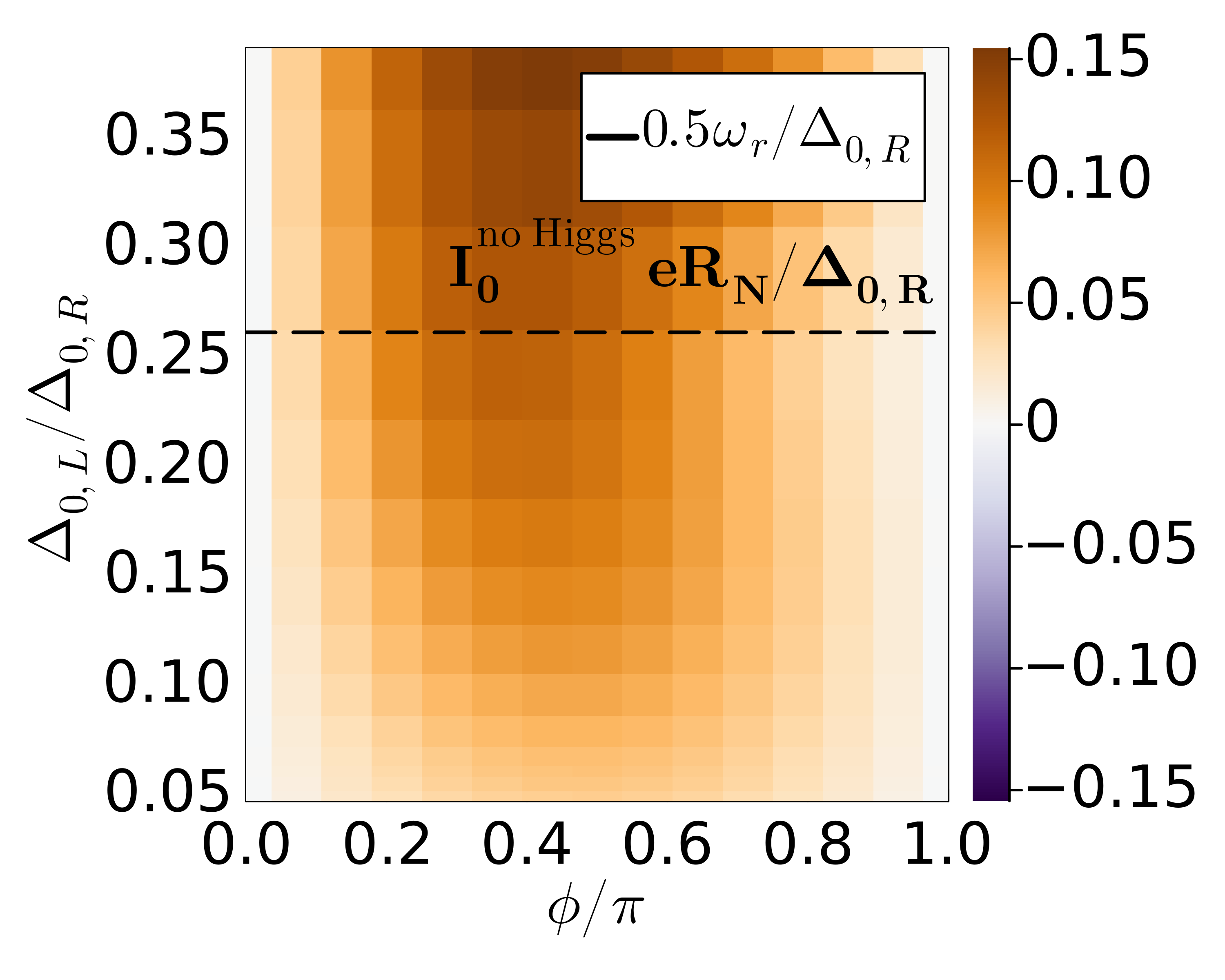}
\end{subfigure}
\begin{subfigure}[b]{0.3295\linewidth}
\caption{}\label{subfig:1cprd}
\includegraphics[width=\columnwidth]{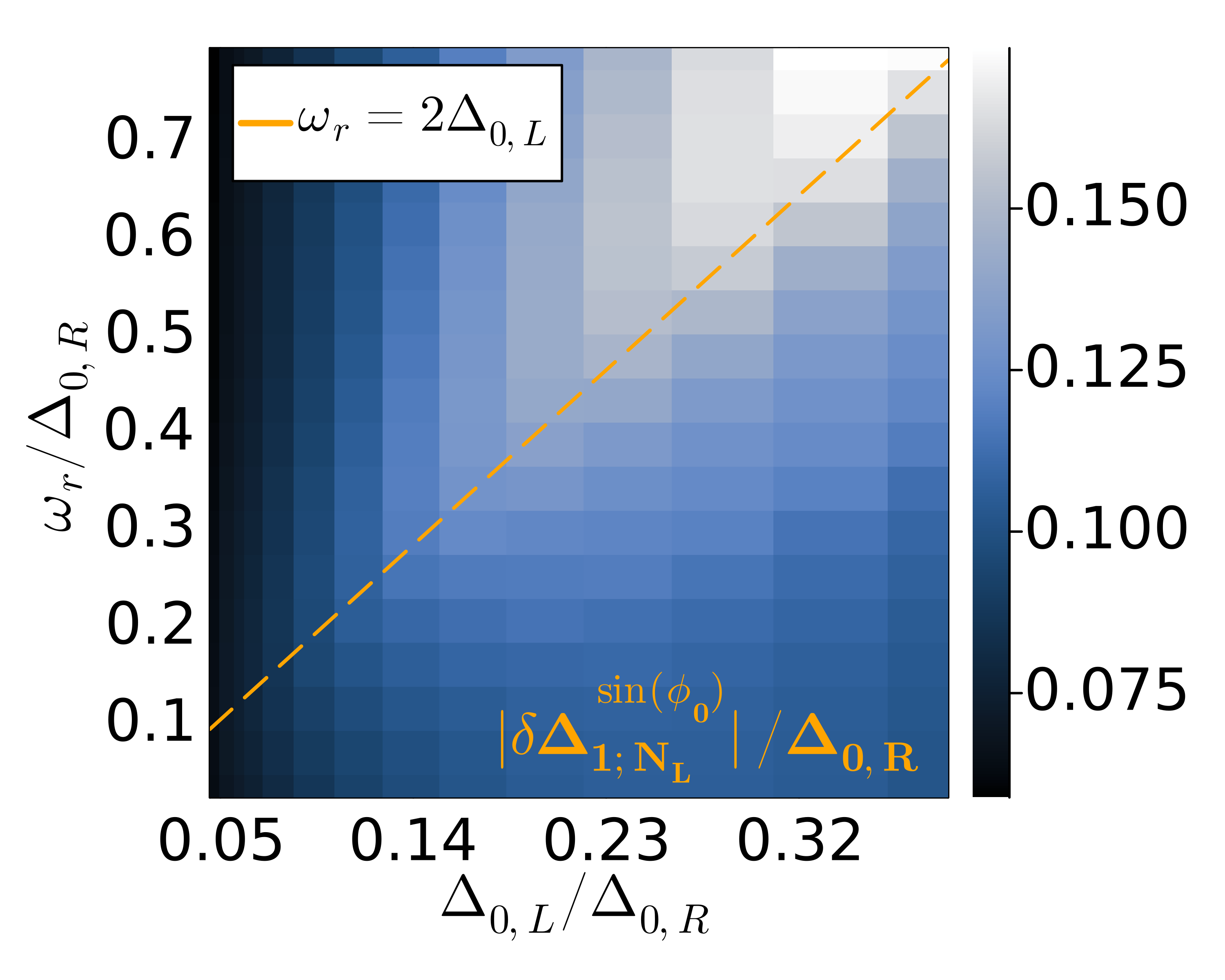}
\end{subfigure}
\begin{subfigure}[b]{0.3295\linewidth}
\caption{}\label{subfig:1cpre}
\includegraphics[width=\columnwidth]{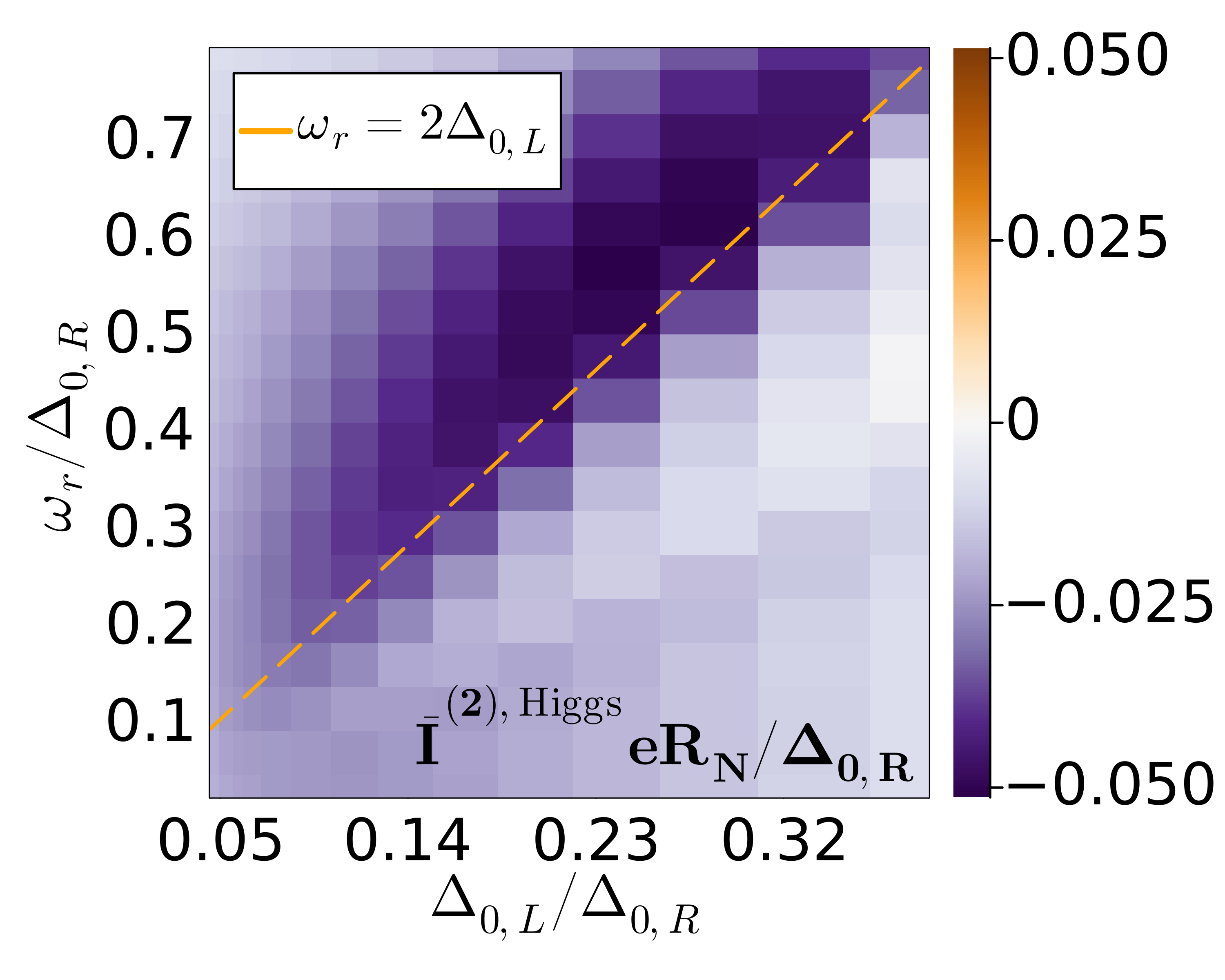}
\end{subfigure}
\begin{subfigure}[b]{0.3295\linewidth}
\caption{}\label{subfig:1cprf}
\includegraphics[width=\columnwidth]{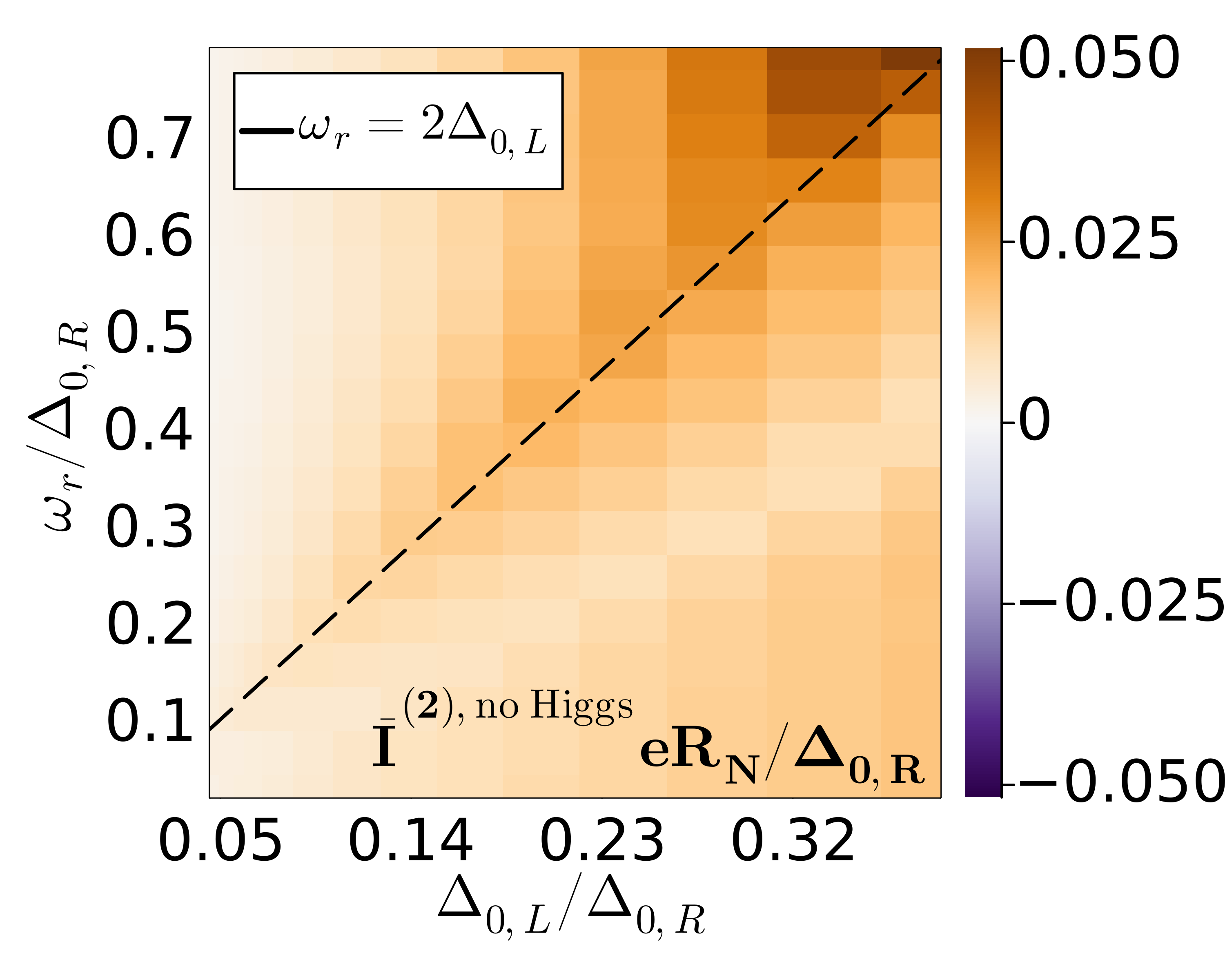}
\end{subfigure}
\caption{(a--d) Numerically obtained CPR ($I_0$) at $\alpha=2$, in a system with $N_L=18$, $N_R=5$~\cite{sizejust}, $\mathcal{T}/\zeta=0.4$ (transparency $\approx 0.48$~\cite{Cuevas2002}), and $\Gamma=0.0125\Delta_{0,R}=0.0025\zeta$. We present the normalized quantity $I_0 e R_N / \Delta_{0,R}$, where $R_N$ is the numerically obtained normal-state resistance. (a--b) CPR as a function of the equilibrium gap asymmetry $\Delta_{0,L}/\Delta_{0,R}$. Each panel considers a different $\omega_r$. The black dashed line marks the point $0.5\omega_r/\Delta_{0,R}=\Delta_{0,L}/\Delta_{0,R}$, corresponding to the resonance condition $\omega_r=2\Delta_{0,L}$. The Higgs renormalization weakens as $2\Delta_{0,L}$ exceeds $\omega_r$, restoring the conventional $\sim\sin(\phi_0)$ behavior. In contrast, when $2\Delta_{0,L}$ moves below $\omega_r$, the CPR develops a negative dip near $\phi_0=0$, originating from the Higgs-induced term $\bar{I}^{(2),\text{Higgs}}\sin(2\phi_0)$ with $\bar{I}^{(2),\text{Higgs}}<0$ [see Fig.~\ref{Fig2}(b) for the sign at $\alpha=2$]. Panels (c–d) show the corresponding results without Higgs renormalization. We do not find any negative dips in $I_0$. (e) The component of the OP varying as $\sin(\phi_0)$ (the $\cos(\phi_0)$ component is negligible) in the left lead at the junction ($x=N_L$), which is peaked at the Higgs resonance $\omega_r=2\Delta_{0,L}$. This resonance is marked by a dashed line in all remaining panels. (f) Second harmonic $\bar{I}^{(2),\text{Higgs}}$ as a function of $\omega_r$ and $\Delta_{0,L} / \Delta_{0,R}$, showing the Higgs resonance. The Higgs-free counterpart in panel (g) exhibits the opposite sign [see Fig.~\ref{Fig2}(b)]. In this case, the peak at $\omega_r=2\Delta_{0,L}$ corresponds to the pair-breaking threshold. }
\label{Fig3} 
\end{figure*}

\begin{figure*}[!htb]
\begin{minipage}[t]{0.3295\linewidth}
\begin{subfigure}[b]{\linewidth}
\caption{}\label{subfig:1aa}
\includegraphics[height=1.99in]{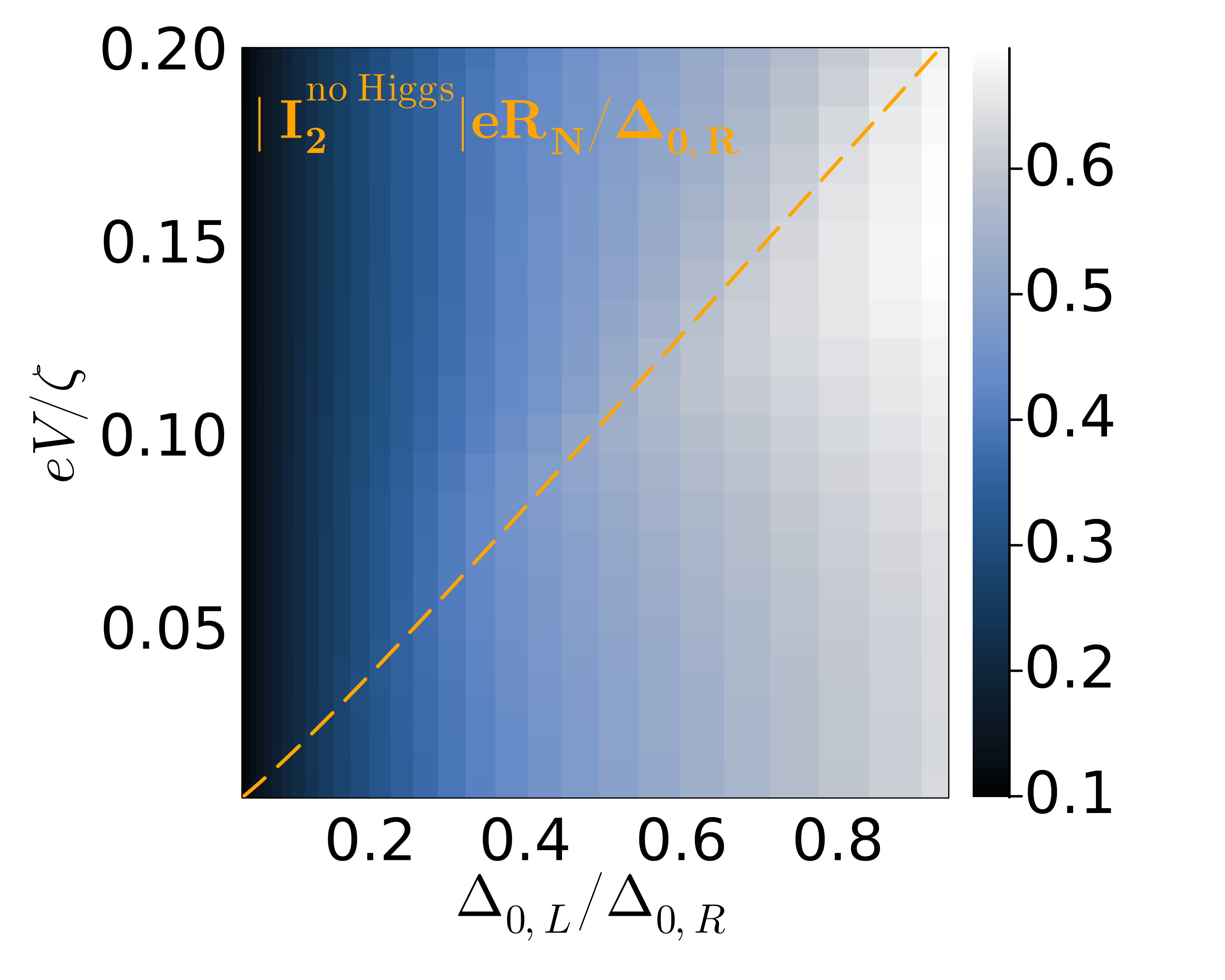}
\end{subfigure}
\begin{subfigure}[b]{\linewidth}
\caption{}\label{subfig:1ad}
\includegraphics[height=1.99in]{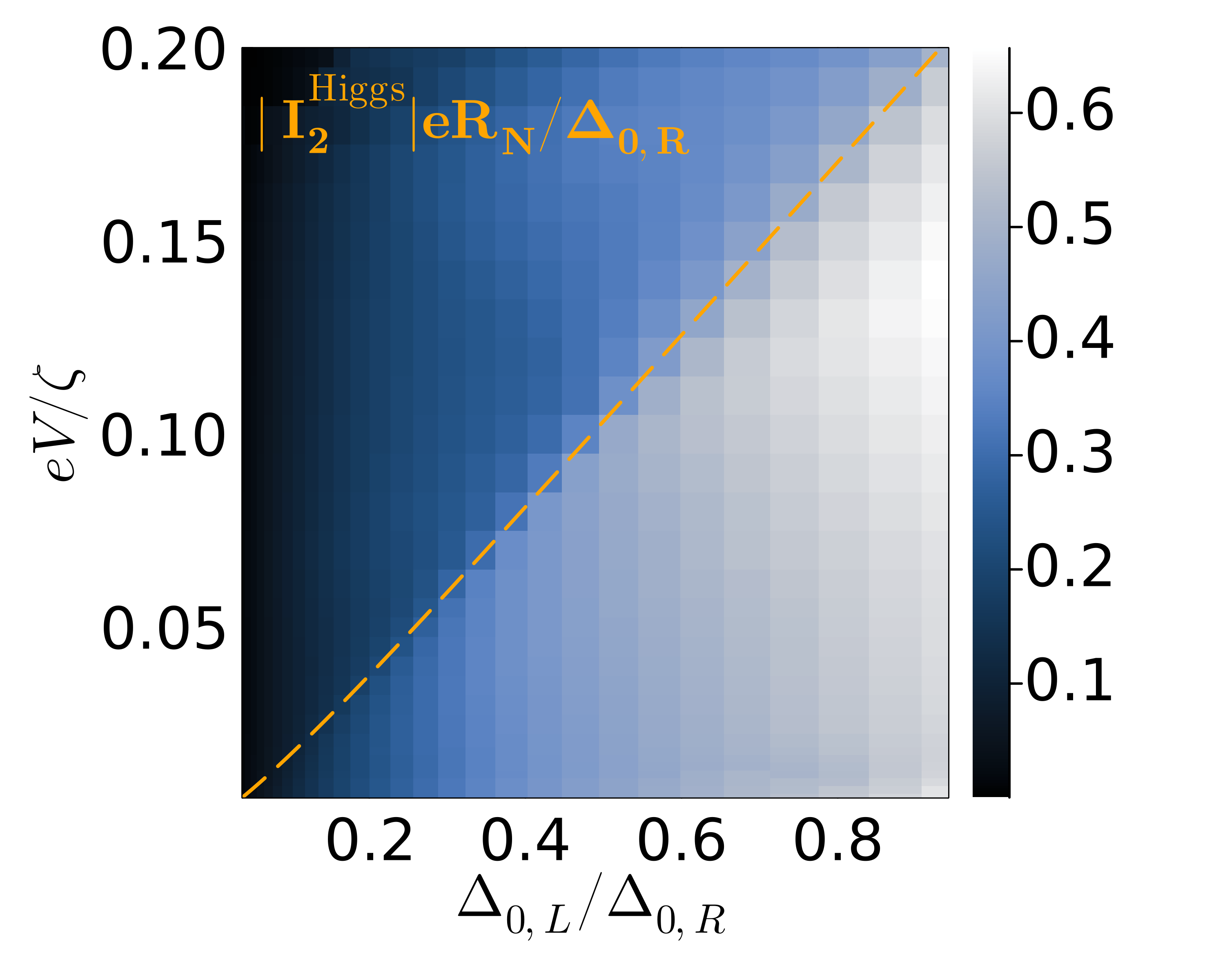}
\end{subfigure}
\end{minipage}
\hfill
\begin{minipage}[t]{0.3295\linewidth}
\begin{subfigure}[b]{\linewidth}
\caption{}\label{subfig:1ac}
\includegraphics[height=1.99in]{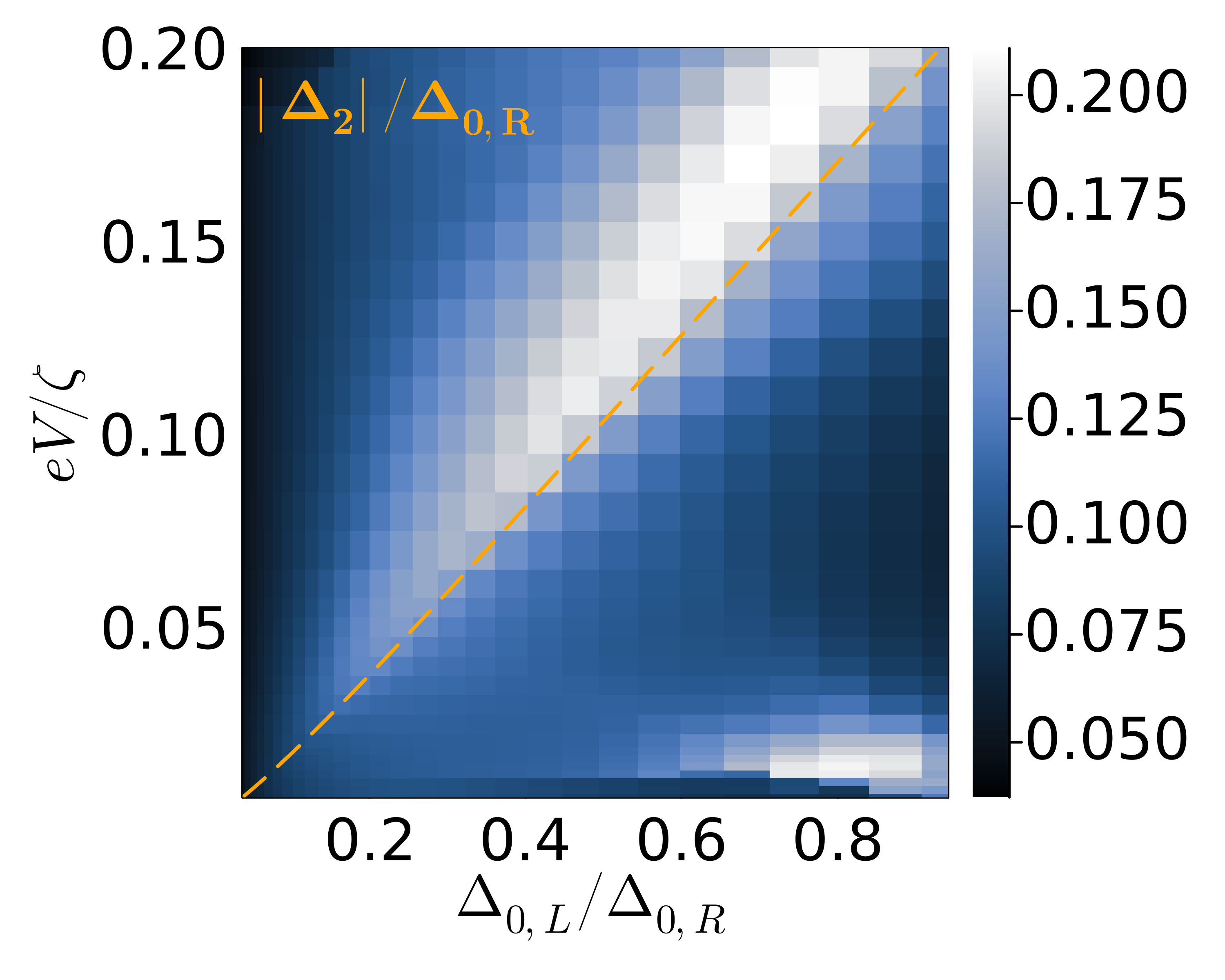}
\end{subfigure}
\begin{subfigure}[b]{\linewidth}
\caption{}\label{subfig:1af}
\includegraphics[height=1.99in]{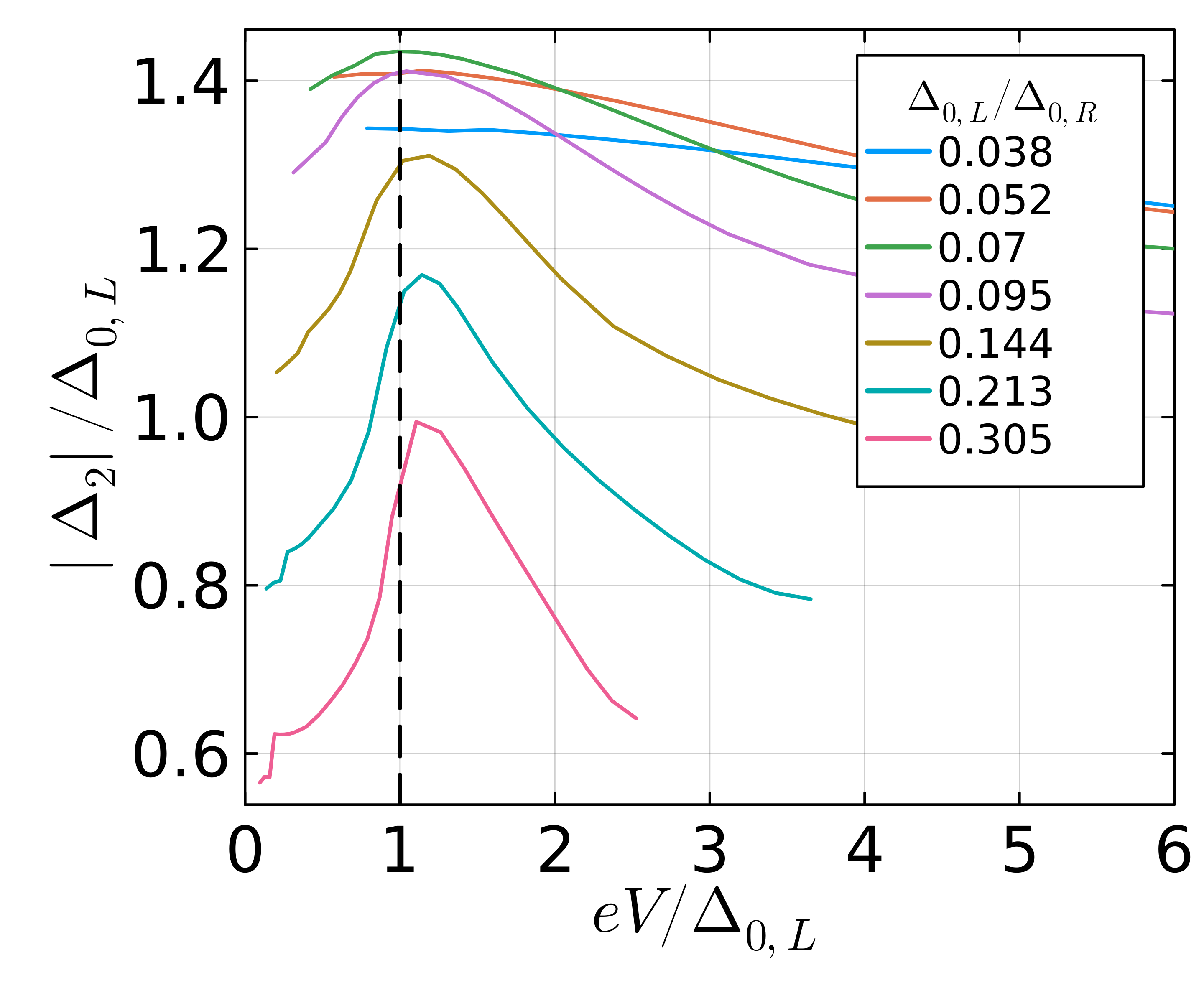}
\end{subfigure}
\end{minipage}
\hfill
\begin{minipage}[t]{0.3295\linewidth}
\begin{subfigure}[b]{\linewidth}
\caption{}\label{subfig:1ab}
\begin{overpic}[height=1.99in]{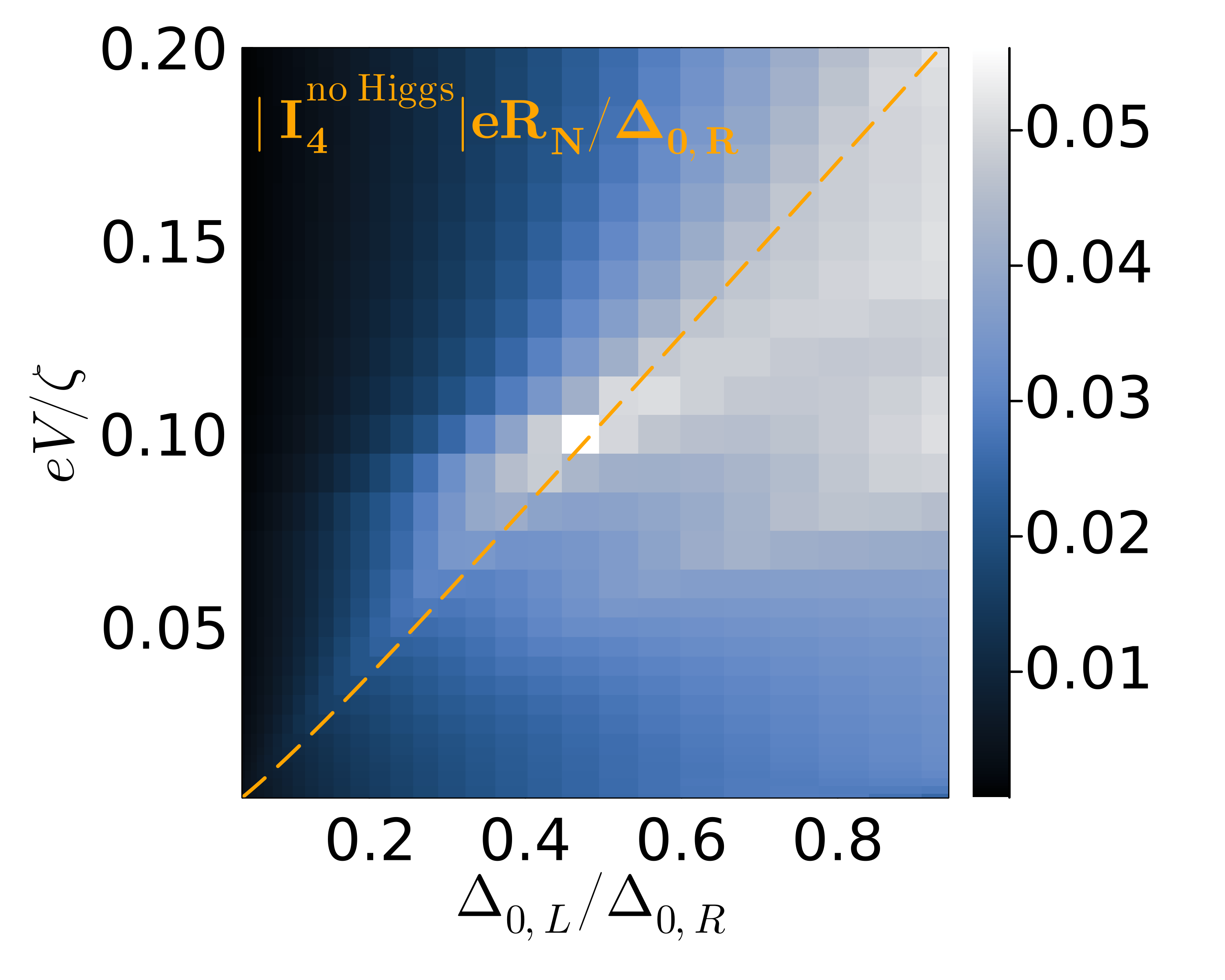}
  \put(19.5,14.8){%
    \begin{tikzpicture}
      \draw[line width=1pt,red] (1,1) rectangle (2.5,2.5);
      \fill[red,opacity=0.05] (1,1) rectangle (2.5,2.5);
    \end{tikzpicture}
  }
\end{overpic}
\end{subfigure}
\begin{subfigure}[b]{\linewidth}
\caption{}\label{subfig:1ae}
\begin{overpic}[height=1.99in]{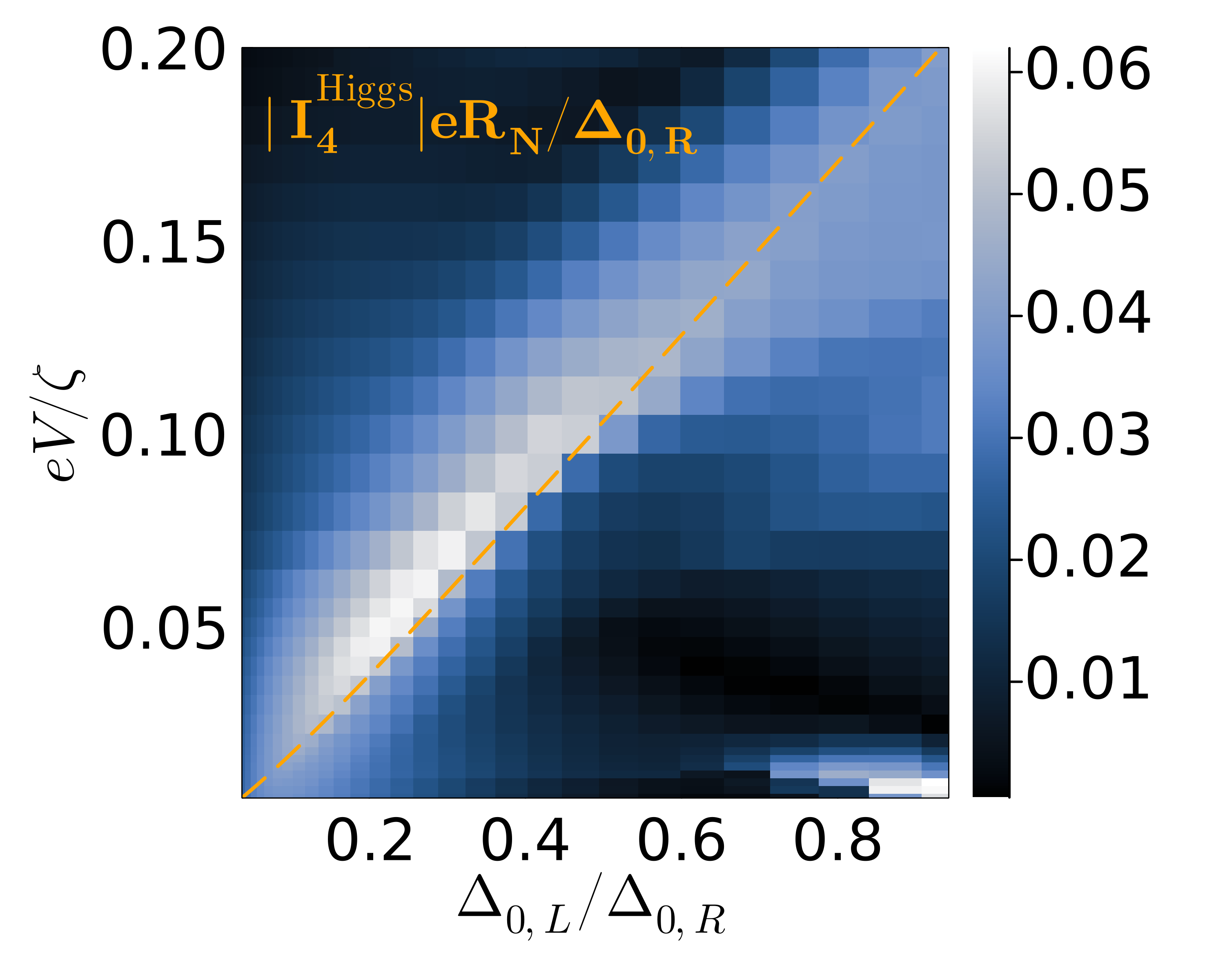}
  \put(19.5,14.8){%
    \begin{tikzpicture}
      \draw[line width=1pt,red] (1,1) rectangle (2.5,2.5);
      \fill[red,opacity=0.05] (1,1) rectangle (2.5,2.5);
    \end{tikzpicture}
  }
\end{overpic}
\end{subfigure}
\end{minipage}
\caption{Numerically obtained current harmonics without radiation, in a system with $N_L=18$, $N_R=5$~\cite{sizejust}, $\mathcal{T}/\zeta=0.4$ (transparency $\approx 0.48$~\cite{Cuevas2002}), and $\Gamma=0.0125\Delta_{0,R}=0.0025\zeta$. Panels (a) and (b) show $I_{\omega_J}$ without ($I^{\text{no Higgs}}_{2}$) and with ($I_{2}^{\text{Higgs}}$) Higgs oscillations, respectively; they exhibit no significant qualitative differences. The orange diagonal dashed line marks the Higgs resonance, $2eV=2\Delta_{0,L}$, in all the heat maps. (c) $\Delta_2$ (OP component oscillating at frequency $\omega_J$), normalized by $\Delta_{0,R}$, which exhibits a peak at the Higgs resonance (orange dashed line). The small offset from the resonance likely originates from a local enhancement of $\Delta_{0,L}$ near the junction barrier due to the proximity effect. (d) Cuts of $\Delta_2$, normalized by $\Delta_{0,L}$, as a function of $eV$ for various values of $\Delta_{0,L}$. It is peaked at the Higgs resonance (vertical dashed line), with the peak achieving its maximum value for $\Delta_{0,L}/\Delta_{0,R}\approx 0.07$. Panels (e) and (f) show $|I_{2\omega_J}|$, in the absence ($I^{\text{no Higgs}}_{4}$) and presence ($I_{4}^{\text{Higgs}}$) of Higgs renormalization, respectively. In the absence of Higgs renormalization (e), only a small $2\omega_J$ current appears, attributable to higher-order Josephson effect. In contrast, with Higgs renormalization (f), a pronounced peak emerges at the Higgs resonance (orange dashed line), particularly in the highly asymmetric regime where $\Delta_{0,L} \ll \Delta_{0,R}$ (bottom-left corner, highlighted by the red box). In the same regime (once again, highlighted by a red box), panel (e) shows that $I^{\text{no Higgs}}_{4}$ remains much smaller.  }
\label{Fig4} 
\end{figure*}

\section{Results}
\label{resul}
In this section, we present our numerical results for the CPR, and the SSs, following the theory developed in the previous sections. We highlight the signatures of the Higgs mode in each case, contrasting with the corresponding results in the absence of the Higgs mode. 

Throughout this work, we use a system of size $N_L=18$, $N_R=5$, $\mathcal{T}/\zeta=0.4$ corresponding to the transparency $\approx0.48$~\cite{Cuevas2002}, $\Gamma/\Delta_{0,R}=0.0125$, along a BCS coupling constant which leads to $\Delta_{0,R}/\zeta=0.2$. The remaining parameters specific to each figure are specified in the corresponding captions. We discuss further practical considerations concerning the essential parameters below in Sec.~\ref{discconc}.

\subsection{Phase bias: Current-phase relation}
\label{resulcpr}
We start by exploring the first proposal in Fig.~\ref{Fig2}, presenting the Higgs renormalized CPR $I_0$ (DC supercurrent) in the absence of a DC voltage bias. We verify key predictions from the phenomenological analysis regarding how both the CPR and the OP depend on the radiation strength $\alpha$, keeping the radiation frequency $\omega_r$ fixed. We begin by examining the Higgs-induced OP modulation, as it ultimately determines how the current depends on $\alpha$. Specifically, Fig.~\ref{Fig2}(a) confirms the prediction of Eq.~\eqref{deltaDeltazerov}: the first Floquet component of the left order parameter at the junction, $\delta \Delta_{1;N_L}$, varies with the Josephson phase as $\sin(\phi_0)$ and scales with $\alpha$ as $J_1(\alpha)$. We find that the $\cos(\phi_0)$ component is negligible. Next, before addressing the full CPR, we first examine the behaviour of its second-harmonic component, which varies with the Josephson phase as $\sim \sin(2\phi_0)$ [cf.~Eq.~\eqref{IDCAA}]. Its amplitude is calculated as $\bar{I}^{(2)}=(2/\pi)\int_0^\pi d\phi_0 I_0(\phi_0) \sin(2\phi_0)$ where $I_0$ is numerically obtained using Eq.~\eqref{Ifcpr}. Fig.~\ref{Fig2}(b) shows the qualitative difference in $I^{(2)}$ with and without Higgs renormalization: in the Higgs-free case ($\bar{I}^{(2),\text{no Higgs}}$) it follows the expected $J_0(2\alpha)$ scaling [cf.~Eq.~\eqref{IDCAA}], is negative near $\alpha=0$, and remains small, whereas including the Higgs renormalization strongly enhances $I^{(2),\text{Higgs}}$, and crucially, makes it positive at $\alpha=0$. Note that we enforce the absence of Higgs oscillations by using a static gap, retaining only the zeroth Floquet component while self-consistently solving for the gap using Eq~\eqref{gapeq}. We reiterate that with Higgs renormalization, the second-harmonic component not only intensifies in amplitude but it also exhibits a distinct dependence on the radiation strength [compare Eqs.~\eqref{IDCAA} and~\eqref{Iup2Higgs}]. Indeed, as shown in Fig.~\ref{Fig2}(b), the Higgs renormalized second harmonic is well captured by the fit $\sum_m p_m[J_m(\alpha)]^2$, as obtained in Eq.~\eqref{Iup2Higgs}. Finally, we are now able to understand the full CPR shown in Fig.~\ref{Fig2}(c) and (d), whose distinctive behaviour in the presence of the Higgs renormalization derives from that of its second harmonic component, $I^{(2),\text{Higgs}}$. In the Higgs-free case [Fig.~\ref{Fig2}(c)], noting that $\bar{I}^{(2),\text{no Higgs}}\ll \bar{I}^{(1)}$ for the transparencies and equilibrium gap asymmetries considered in this work, we obtain from Eq.~\eqref{IDCAA} $I_0\approx \bar{I}^{(1)}\sin(\phi_0)$. Thus, the CPR maintains its typical $\sin(\phi_0)$ phase dependence, along with the associated $\sim J_0(\alpha)$ $\alpha-$dependence. In contrast, in the presence of Higgs renormalization [Fig.~\ref{Fig2}(d)], the CPR is strongly modified by the enhanced second harmonic $\bar{I}^{(2),\text{Higgs}}$, with $I_0\approx \bar{I}^{(1)}\sin(\phi_0)+\bar{I}^{(2),\text{Higgs}}\sin(2\phi_0)$. This not only introduces a clear $\sin(2\phi_0)$ phase dependence, but also leads to a distinct $\alpha$ dependence via $\bar{I}^{(2),\text{Higgs}}$ consistent with Eq.~\eqref{Iup2Higgs}. We emphasize that even if $\bar{I}^{(2),\text{no Higgs}}$ were sizable, it may be clearly distinguished from the Higgs-renormalized case as their signs and $\alpha$ dependences differ [Fig.~\ref{Fig2}(b)], both of which are reflected in the full CPR. When the second harmonic is negligible, the CPR [$\sim\sin(\phi_0)$] is symmetric about $\phi_0 = \pi/2$ and remains positive for $\phi_0 \in (0, \pi)$. However, a negative (positive) second harmonic skews the CPR backward (forward). Moreover, if the second harmonic is sufficiently large, $\lvert \bar{I}^{(2),\text{Higgs}} \rvert > 0.5\bar{I}^{(1)}$, it can drive the CPR below zero near $\phi_0 = 0$ ($\phi_0 = \pi$).

Next, in Fig.~\ref{Fig3} we complement the analysis in Fig.~\ref{Fig2} by studying the dependence of the CPR and OP on $\omega_r$ and equilibrium gap asymmetry, for a fixed value of $\alpha$. In Fig.~\ref{Fig3}(a--b), we plot the CPR as a function of the equilibrium gap asymmetry for two fixed values of $\omega_r$, at a fixed radiation strength $\alpha=2$. A clear transition appears at $\omega_r = \omega_{H,L}=2\Delta_{0,L}$; for $\omega_r>2\Delta_{0,L}$, the CPR dips below zero near $\phi_0=0$, signaling a sizeable negative Higgs-induced $\bar{I}^{(2),\text{Higgs}}$ [see Fig.~\ref{Fig2}(b), which shows $\bar{I}^{(2),\text{Higgs}}<0$ at $\alpha=2$ when $\omega_r$ is slightly above $2\Delta_{0,L}$]. Instead, for $\omega_r<2\Delta_{0,L}$, this negative dip disappears. The corresponding plots in the absence of Higgs renormalization, shown in Figs.~\ref{Fig3}(c) and (d), do not exhibit such a transition, lacking a negative dip altogether. This suggests $\bar{I}^{(2),\text{no Higgs}}$ is negligible. In Fig.~\ref{Fig3}(e), we see the resonant enhancement of the OP at $\omega_r=\omega_{H,L}$, as predicted by Eq.~\eqref{deltaDeltazerov}. Recall from the same equation that the harmonics $\delta\Delta_m$ decay as $J_m(\alpha)$. No signatures of higher-order resonances at $m\omega_r = \omega_{H,L}$ are observed for $\alpha=2$. In Figs.~\ref{Fig3}(f) and (g), we show the amplitudes of the second harmonic component with and without Higgs renormalization, respectively. The former is negative, and shows a resonant enhancement at $\omega_r=\omega_{H,L}$, while the latter is positive [see Fig.~\ref{Fig2}(b)].

\begin{figure*}[!htb]
\begin{subfigure}[b]{0.49\linewidth}
\caption{}\label{subfig:1a}
\includegraphics[height=2.45in]{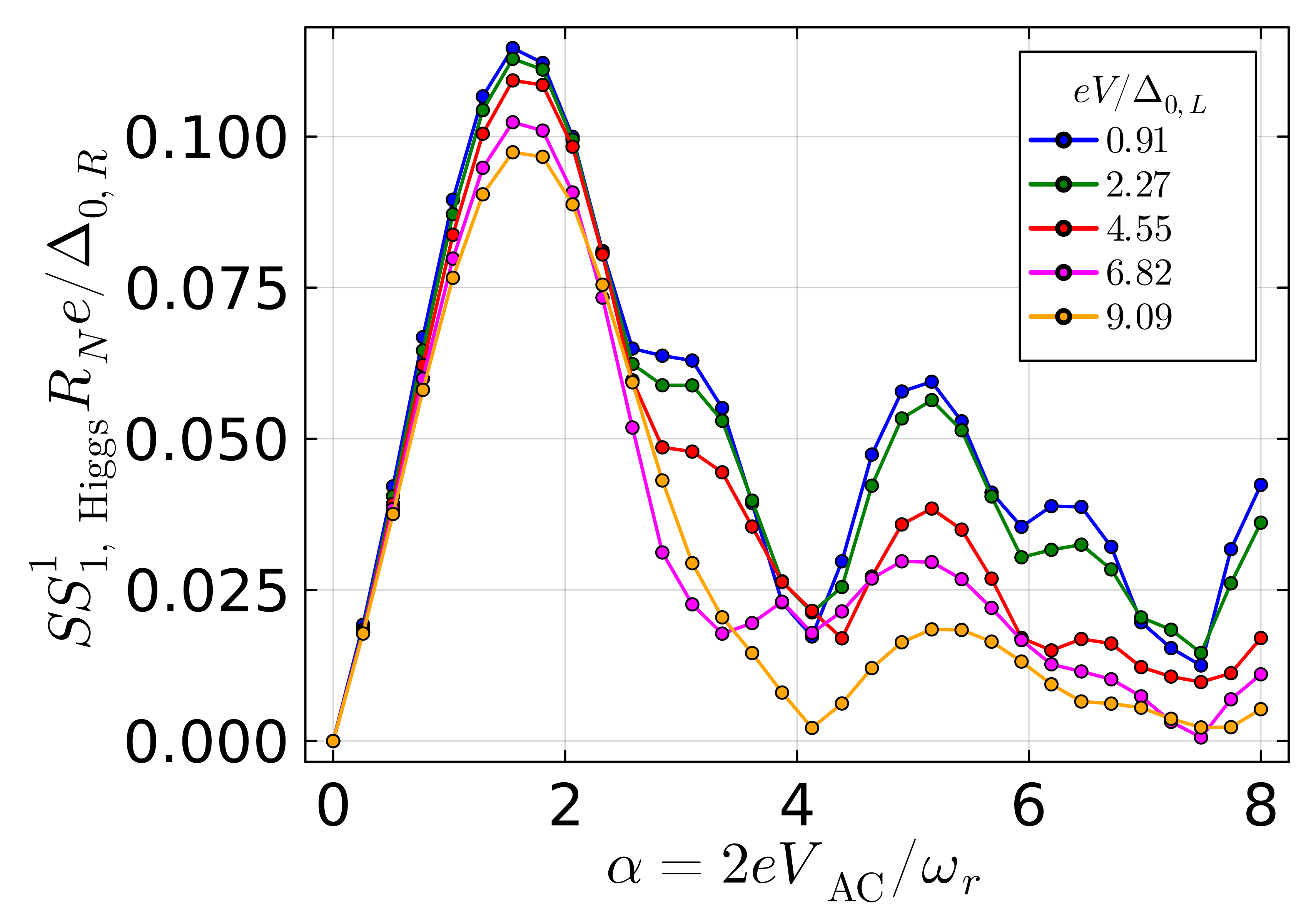}
\end{subfigure}
\begin{subfigure}[b]{0.49\linewidth}
\caption{}\label{subfig:1b}
\includegraphics[height=2.45in]{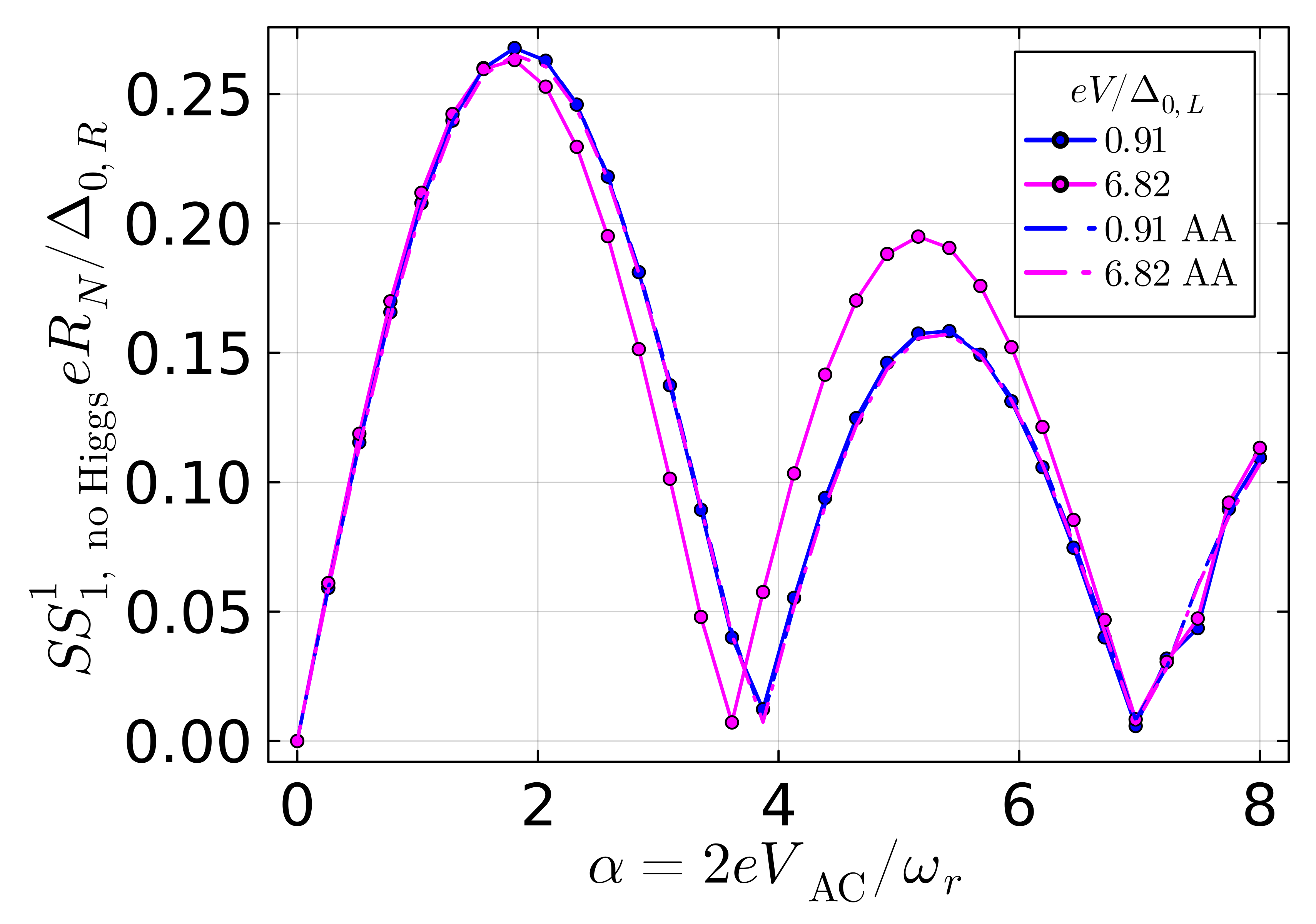}
\end{subfigure}
\begin{subfigure}[b]{0.49\linewidth}
\caption{}\label{subfig:1c}
\includegraphics[height=2.45in]{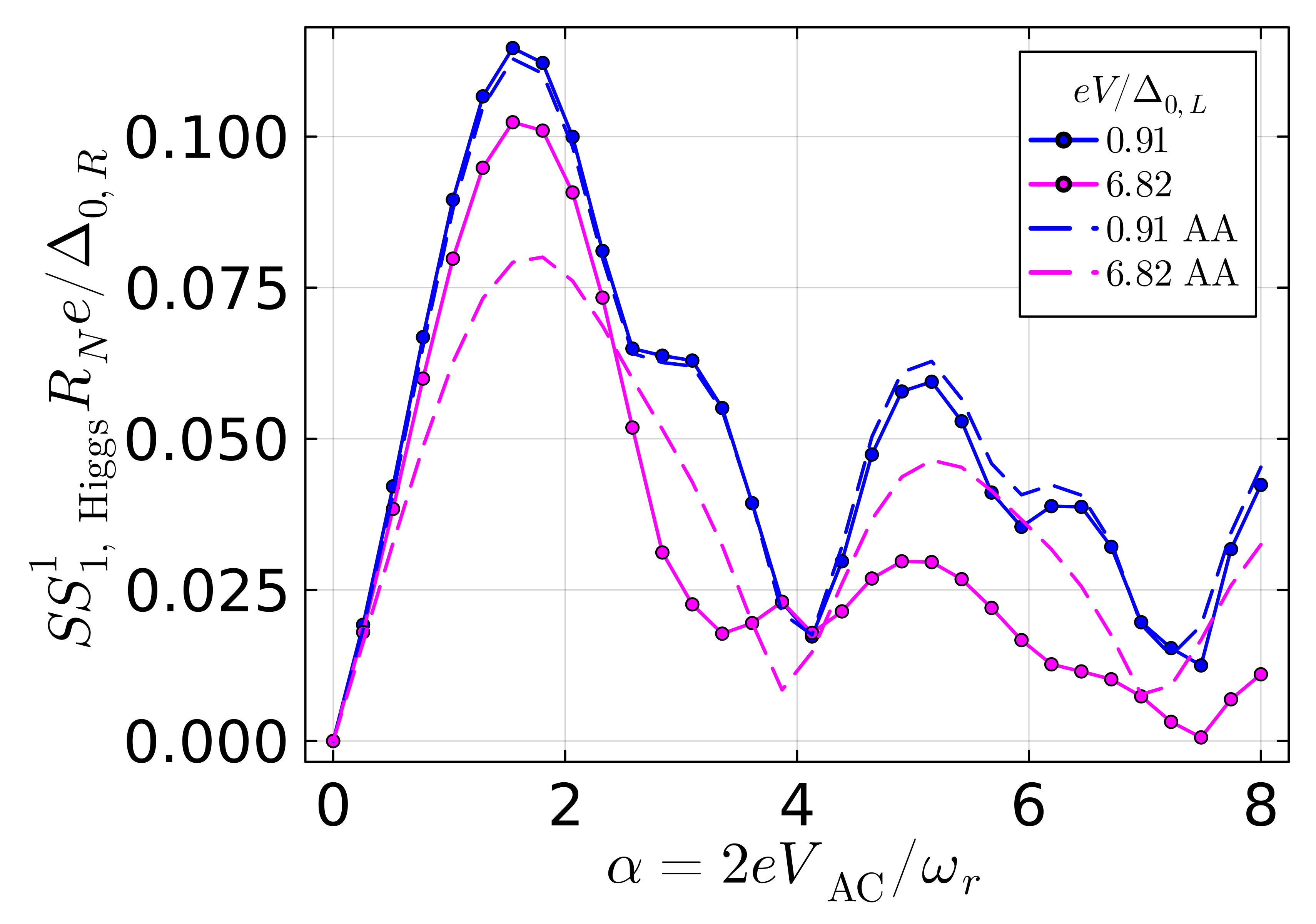}
\end{subfigure}
\begin{subfigure}[b]{0.49\linewidth}
\caption{}\label{subfig:1d}
\includegraphics[height=2.45in]{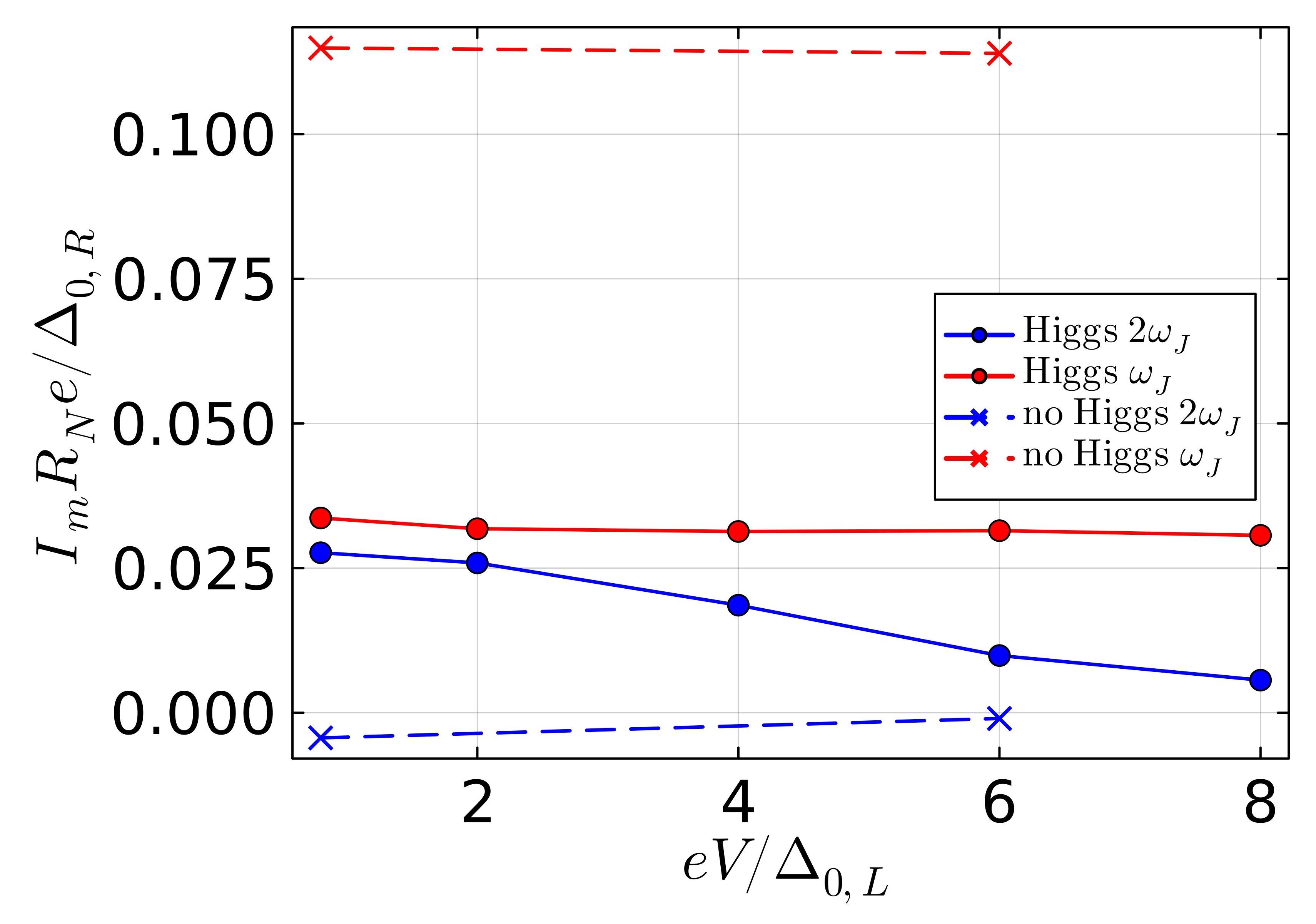}
\end{subfigure}
\caption{Numerically calculated Shapiro step height $SS^1_1$ [Eq.~\eqref{SSab}], in a system with $N_L=18$, $N_R=5$~\cite{sizejust}, $\mathcal{T}/\zeta = 0.4$ (corresponding to transparency$\approx 0.48$~\cite{Cuevas2002}), $\Delta_{0,L} = 0.045\Delta_{0,R} = 0.0088\zeta$, and $\Gamma = 0.0125\Delta_{0,R} = 0.0025\zeta$. We plot the normalized quantity $SS^1_1 R_N e / \Delta_{0,R}$ as a function of the radiation strength $\alpha$, for various values of the DC voltage $eV$ sweeping across and beyond the Higgs resonance $eV=\Delta_{0,L}$. We show it in (a) the presence and (b) the absence of Higgs renormalization. In its presence, $SS^1_{1,\text{Higgs}}$ noticeably deviates from the $J_{-1}(\alpha)$ profile, evident most immediately from the changes in the location and magnitudes of the nodes/dips as a function of $\alpha$. On the other hand, in the absence of Higgs renormalization [panel (b)], while $SS^1_{1,\text{no Higgs}}$ is still altered by the presence of $I_{2\omega_J}$ arising solely from higher-order Josephson processes [cf. Eq.~\eqref{SS11n}], for the chosen transparency this contribution is insufficient to generate noticeable deviations from the expected $\sim J_{-1}(\alpha)$ dependence. Panel (c) compares the exact numerical results from (a) (solid lines with markers) with the AA of Eq.~\eqref{HiggsSS11} (dashed lines). The AA is evaluated using the numerically obtained currents [Eq.~\eqref{If}] with $I_{\omega_J} = \Im(I_{20} - I_{-20})$ and $I_{2\omega_J} = \Im(I_{40} - I_{-40})$. Our goal is to test whether the $\alpha$-dependence of $S^1_1$ (but not that on the DC voltage) can be captured by the AA [Eq.~\eqref{SS11n}]. For small $\omega_J$ and $\omega_r$, the AA agrees well with the numerical results, reproducing the Higgs-induced deviations from the $J_{-1}(\alpha)$ profile shown in (b). Finally, panel (d) presents the amplitudes of the $\omega_J$ and $2\omega_J$ currents, with $I_{\omega_J} = I_2$ (red) and $I_{2\omega_J} = I_4$ (blue), both with and without Higgs renormalization, as functions of the bias voltage. $I_4$ decreases away from the resonance.}
\label{Fig5} 
\end{figure*}

\subsection{Voltage bias: Shapiro steps}
\label{resulss}
Let us now discuss the signature of the Higgs mode in the Shapiro steps that appear as a consequence of the phase locking 
between the microwave field and the applied DC voltage bias. In Fig.~\ref{Fig4}, to clarify the impact of Higgs renormalization, we proceed by examining the OP and the current harmonics $I_{\omega_J}$ and $I_{2\omega_J}$ in DC voltage biased JJs in the absence of radiation, comparing results with and without the Higgs renormalization. We find that $I_{\omega_J}$ exhibits no significant qualitative changes in the presence of Higgs renormalization [compare Figs.~\ref{Fig4}(a) and (d)]. The Higgs oscillations mainly appear as a nonequilibrium OP component oscillating at frequency $\omega_J$, which is shown in Figs.~\ref{Fig4}(c) and (d). Its amplitude peaks at the Higgs resonance, $\omega_J=\omega_{H,L}$. In highly asymmetric junctions, these Higgs oscillations result in a pronounced peak in the second harmonic Josephson current $I_{2\omega_J}^{\text{Higgs}}$ as the bias voltage is tuned across the Higgs resonance [Fig.~\ref{Fig4}(f)]. By contrast, without Higgs oscillations, $I_{2\omega_J}^{\text{no Higgs}}$ is much smaller and no such peak arises [Fig.~\ref{Fig4}(e)].

\begin{figure*}[!htb]
\begin{subfigure}[b]{0.49\linewidth}
\caption{}\label{subfig:2a}
\includegraphics[height=2.45in]{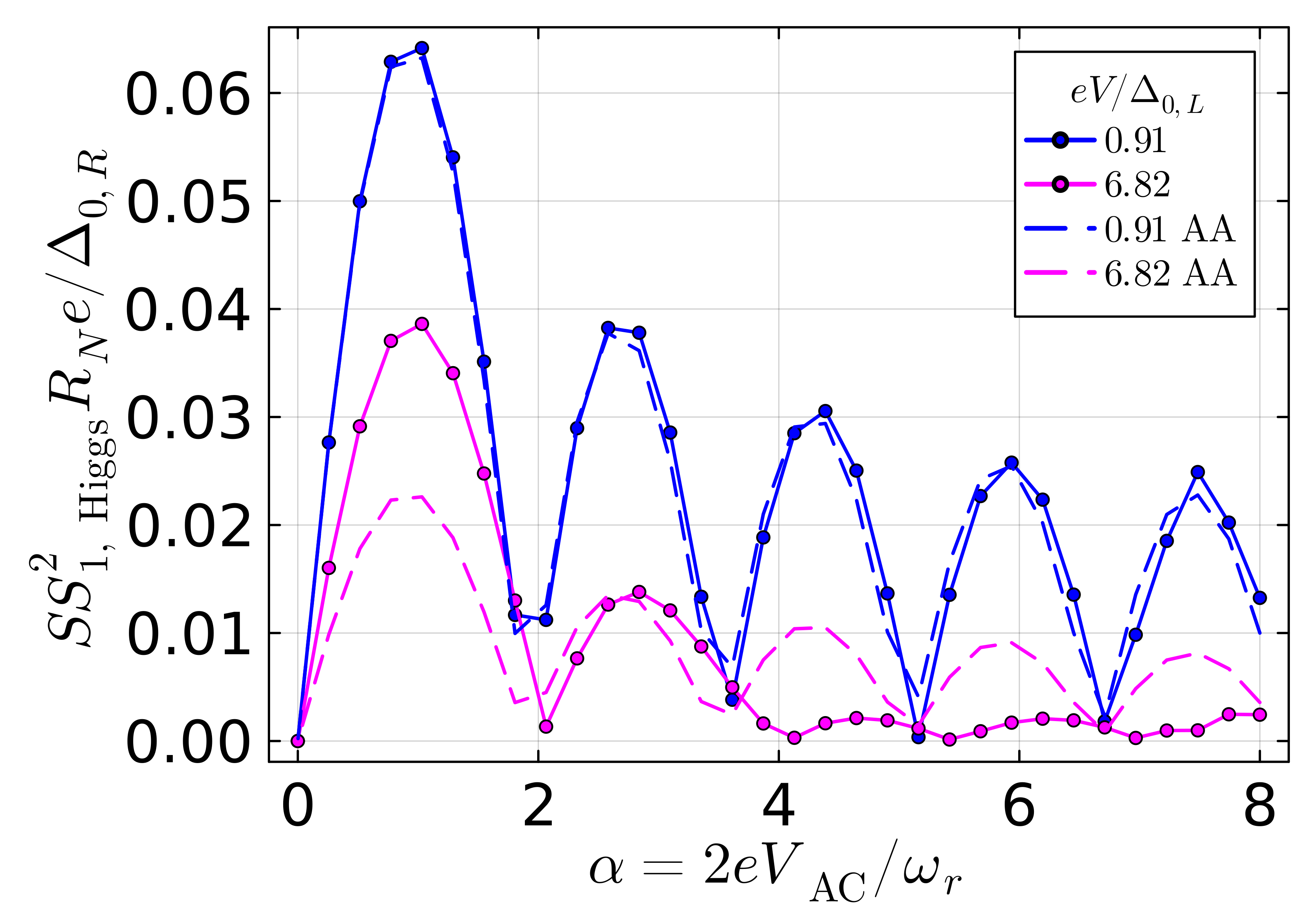}
\end{subfigure}
\begin{subfigure}[b]{0.49\linewidth}
\caption{}\label{subfig:2b}
\includegraphics[height=2.45in]{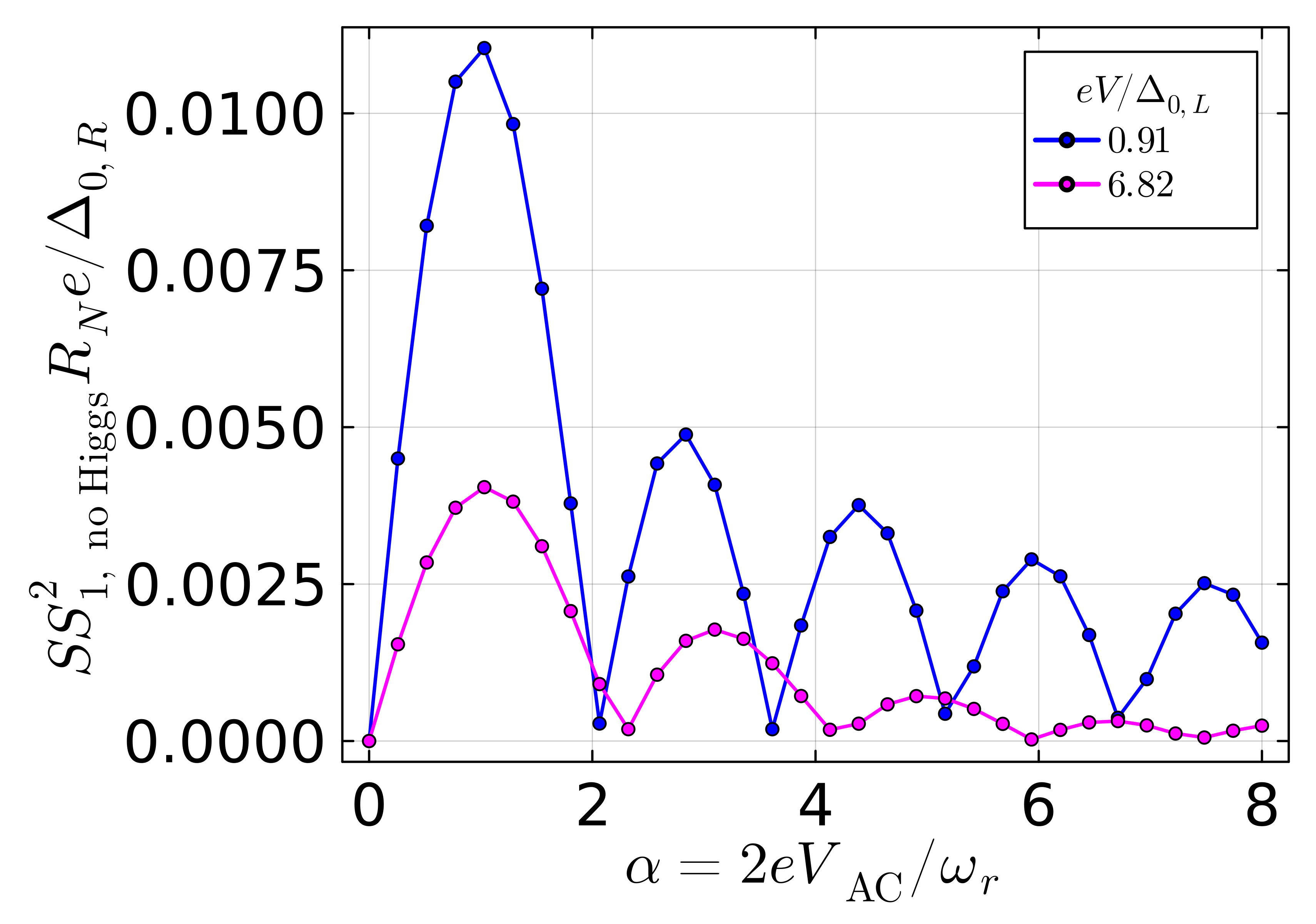}
\end{subfigure}
\caption{(a) Same as Fig.~\ref{Fig5}(c) (including the parameters, except for $\omega_r$), but we show $SS^2_1$, where $2\omega_J=\omega_r$. $SS^2_{1,\text{Higgs}}$, which arises from the Higgs enhanced $I_{2\omega_J}^{\text{Higgs}}$, is approximately of the same magnitude as $SS^1_{1,\text{Higgs}}$ in Fig.~\ref{Fig5} as $I_{2\omega_J}^{\text{Higgs}}\sim I_{\omega_J}^{\text{Higgs}}$. Similar to Fig.~\ref{Fig5}(c), we overlay the AA results given by Eq.~\eqref{HiggsSS21} using dashed lines. These match the numerical results only for small $\omega_J$ and $\omega_r$. (b) Same as (a), but we show the data in the absence of Higgs renormalization. Notice that now $SS^2_{1,\text{no Higgs}}$ is about an order of magnitude smaller than its Higgs renormalised counterpart in (a). For the chosen value of transparency $\approx0.48$, the higher-order Josephson current $I^{\text{no Higgs}}_{2\omega_J}$ is far from sufficient to match the Higgs renormalised $I_{2\omega_J}^{\text{Higgs}}$ [c.f. Eq.~\eqref{SS21}. See also Fig.~\ref{Fig5}(d)]. }
\label{Fig6} 
\end{figure*}

Now, armed with an understanding of the effect of the Higgs renormalization on the AC Josephson current harmonics, we are ready to look at how these manifest in the SSs in irradiated DC voltage biased JJs in Figs.~\ref{Fig5} and~\ref{Fig6}. Specifically, we focus on two SSs, $SS^1_1$ in Fig.~\ref{Fig5} and $SS^2_1$ in Fig.~\ref{Fig6}, which require $\omega_J=\omega_r$ and $2\omega_J=\omega_r$, respectively. Before turning to the role of Higgs renormalization, we first consider the trivial case where it is absent. Even for the high transparencies ($\approx 0.48$) studied here, the $2\omega_J$ current arising from higher-order Josephson processes remains much smaller than the conventional $\omega_J$ current given by Eq.~\eqref{I0}. In fact, as shown in Fig.~\ref{Fig4}(c) and (d), a large equilibrium gap asymmetry ($\Delta_{0,L}\ll\Delta_{0,R}$ without loss of generality) aids the suppression of the $2\omega_J$ current. As such, we expect the current to be described by Eq.~\eqref{I0}, containing only the $\omega_J$ current, $I_{\omega_J}$. Furthermore, while the amplitude of $I_{\omega_J}$ is expected to be dependent on both $\omega_J$ and $\omega_r$ [$I_{\omega_J}(\omega_J,\omega_r)$], for small $\omega_J$, along with radiation of low intensity 
(low $\alpha$) and low frequency $\omega_r$, the frequency dependence of $I_{\omega_J}$ may be neglected and replaced with its 
zero-frequency value. As derived in Eq.~\eqref{SS11}, this yields $SS^1_1=2I_{\omega_J}J_{-1}(\alpha)$ with the distinctive 
$J_{-1}(\alpha)$ profile~\cite{Hamilton1971,Hamilton1972}. For the values of $\alpha$ and $\omega_r$ considered in this work, as 
shown in Fig.~\ref{Fig5}(b), we find that this is indeed the case, with no significant differences from the $J_{-1}(\alpha)$ profile. 
Turning to $SS^2_1$, we recall from Eq.~\eqref{SS21} that it is governed solely by $I_{2\omega_J}$. As noted above, $I_{2\omega_J}^{\text{no Higgs}}$ is much smaller than $I_{\omega_J}^{\text{no Higgs}}$ in the absence of Higgs renormalization. Consequently, $SS^2_{1,\text{no Higgs}}$ has a much smaller magnitude compared to $SS^1_{1,\text{no Higgs}}$, which depends on both $I_{\omega_J}^{\text{no Higgs}}$ and $I_{2\omega_J}^{\text{no Higgs}}´$ [Eq.~\eqref{SS11}]. This difference is clearly visible when comparing their amplitudes in Figs.~\ref{Fig5}(b) and~\ref{Fig6}(b), which reveal that the crests of $SS^2_{1,\text{no Higgs}}$ are nearly two orders of magnitudes smaller than those of $SS^1_{1,\text{no Higgs}}$.

Now, we turn to the case with Higgs renormalization. We begin with $SS^1_{1,\text{Higgs}}$, shown in Fig.~\ref{Fig5}(a), where it is plotted 
as a function of $\alpha$ for several bias voltages $\omega_J$ ranging from below the Higgs resonance ($\omega_J \lesssim \omega_{H,L}$) 
to well above it ($\omega_J \gg \omega_{H,L}$). Recall that $SS^1_{1,\text{Higgs}}$ depends on a combination of $J_{-1}(\alpha)$ and 
$J_{-2}(2\alpha)$, weighted by $I_{\omega_J}^{\text{Higgs}}$ and $I_{2\omega_J}^{\text{Higgs}}$, respectively [cf.\ Eq.~\eqref{HiggsSS21}]. At the resonance $\omega_J\approx \omega_{H,L}$, the strong Higgs renormalized contribution to $I_{2\omega_J}$ amplifies the $J_{-2}(2\alpha)$ 
term, producing clear deviations from the $J_{-1}(\alpha)$ behavior observed in the absence of Higgs renormalization [cf.\ 
Fig.~\ref{Fig4}(d)]. Even more striking is the behavior of $SS^2_1$, which according to Eq.~\eqref{HiggsSS21} depends exclusively 
on $I_{2\omega_J}^{\text{Higgs}}$. Since $I_{2\omega_J}^{\text{Higgs}}$ becomes comparable in size to $I_{\omega_J}^{\text{Higgs}}$ near the resonance, $SS^2_{1,\text{Higgs}}$ attains a magnitude similar to $SS^1_{1,\text{Higgs}}$, as evident in Figs.~\ref{Fig5}(a) and~\ref{Fig6}(a). This provides a clear fingerprint of Higgs renormalization in conventional JJs, as no other mechanism can generate such a pronounced increase in the $2\omega_J$ current.

Now we comment on the applicability of the AA introduced earlier. As shown in Figs.~\ref{Fig5}(c) and~\ref{Fig6}(a), we find that 
for small $\omega_r<\Delta_{0,R}$, our numerical results are matched by the AA (Eqs.~\eqref{HiggsSS11} and~\eqref{HiggsSS21}, 
respectively). Specifically, on starting with the numerically obtained coefficients $I_{\omega_J}=\Im(I_{20}-I_{-20})$ and 
$I_{2\omega_J}=\Im(I_{40}-I_{-40})$ at $\alpha=0$, which represent the amplitudes of the $\sin(\omega_J t)$ and $\sin(2\omega_J t)$ 
currents in the absence of radiation, the numerically obtained $SS^1_1$ and $SS^2_1$ mirror Eqs.~\eqref{HiggsSS11} and~\eqref{HiggsSS21},
 respectively~\cite{nocosI}. For large values of $\omega_r$ comparable to or larger than $\Delta_{0,R}$, we find significant 
 deviations from the AA predictions. The validity of the AA is primarily limited by electronic retardation processes, which 
 imbibe the amplitudes $I_{\omega_J}$ and $I_{2\omega_J}$ with frequency dependence~\cite{Hamilton1971,Hamilton1972}. A 
 detailed account of this argument may be found in Chapter 11 of Refs.~\cite{Likharevbook} and~\cite{Baronebook}. In particular, 
 a leading order perturbative (in the tunnel coupling $\mathcal{T}$) calculation reveals resonances at frequency $\Delta_{0,L/R}$ 
 or $\Delta_{0,L}\pm\Delta_{0,R}$, corresponding to resonant excitation of quasiparticles between the singular gap-edges of 
 the two SC leads. Since the dominant contribution to the current typically arises from the exchange of only the few lowest 
 harmonics of $\omega_J$ and $\omega_r$, for small values of $\omega_r$ and $\omega_J$, the components $I_{\omega_J}$ and 
 $I_{2\omega_J}$ are probed only near zero frequency, and their frequency dependence can therefore be neglected. However, 
 for $\omega_J$ and $\omega_r$ comparable to the resonant frequencies mentioned above, or for larger transparency and $\alpha$ 
 when higher harmonics of $\omega_J$ and $\omega_r$ also participate via higher-order tunneling processes, this frequency 
 dependence may no longer be neglected. It introduces an explicit dependence on the harmonic number in the current amplitudes, 
 which modifies the Bessel function amplitudes derived within the AA.

\section{Discussion and Conclusions}
\label{discconc}
In this work, we demonstrate that the Higgs mode can be excited and detected in highly asymmetric, high-transparency JJs under 
microwave irradiation. We propose two approaches: (i) measuring the second harmonic of the current-phase relation, whose 
Higgs-renormalized dependence on radiation strength deviates strongly from the Higgs-free case, and (ii) applying a DC voltage 
bias, where the Higgs-renormalized second harmonic AC Josephson current can be extracted from Shapiro step measurements.

We note that a recent work by Vallet and Cayssol~\cite{Vallet2024} had explored a similar, yet different scenario, wherein the Higgs mode 
is excited by irradiating the \emph{bulk} of the SC leads by a \emph{uniform} THz electric field. This corresponds to the SC leads being subjected to a uniform bulk vector potential $\mathbf{A}$ oscillating at frequency $\omega_r$. This is not equivalent to our case where 
we assume that the radiation induces a time-dependent potential difference between the two SC leads $V_{AC}(t)$~\cite{Baronebook,Cuevas2002,
Chauvinthesis,Shapiro1963,Kot2020,Siebrecht2023}, with the drop restricted to the barrier region~\cite{Lahiri2025}. 
Indeed, Ref.~\cite{Vallet2024} considers an OP oscillating as $\Delta(t)=\Delta_0+\delta\Delta_{VC}(t)$ where 
$\delta\Delta_{VC}(t)=2\Re(\Delta_2 e^{-i2\omega_r t})$, arising from the second-order coupling of the OP to the oscillating bulk vector
potential~\cite{Tang2020}, which differs from our results as the spectrum of $\delta\Delta_{VC}$ does not contain: (i) any multiples of 
$\omega_J$ and, (ii) the first harmonic of the radiation frequency $\omega_r$. (i) follows from the fact that Ref.~\cite{Vallet2024} 
only considers tunnel JJs; oscillations in $\delta\Delta_{VC}(t)$ at frequency $\omega_J$ are $\mathcal{O}(\mathcal{T}^2)$ and they 
contribute to the current only at $\mathcal{O}(\mathcal{T}^4)$, while the leading order Josephson current is $\mathcal{O}(\mathcal{T}^2)$. 
In our case, by arguments of gauge invariance, it is clear that the OP must respond to the gauge-invariant Josephson phase given by 
Eq.~\eqref{Jphase} and thus, $\delta\Delta(t)$ is bound to contain $\omega_J$, as well as \emph{all} harmonics of $\omega_r$ (not 
just the even ones, or just $2\omega_r$) in its spectrum. Consequently, Ref.~\cite{Vallet2024} does not obtain a Higgs-induced 
correction to $SS^1_1 $ at $\omega_J=\omega_r$, nor do they obtain $SS^2_1 $ at $2\omega_J=\omega_r$, both of which are present in 
our results shown in Figs.~\ref{Fig5} and~\ref{Fig6}. Instead, they only obtain $SS^1_2 $ at $\omega_J=2\omega_R$. Similarly, in the phase-biased case, the resonance at $\omega_r = 2\Delta_{0,L}$ in the second harmonic of the CPR is absent when $\delta\Delta$ lacks a component oscillating at $\omega_r$. Instead, the leading resonance appears at $2\omega_r = 2\Delta_{0,L}$. We remark that these differences result from the two physically distinct excitation processes considered by them and this work.

Finally, we comment on the practical realization of our proposal. Since our aim is to realize highly asymmetric Josephson junctions 
using conventional $s$-wave superconductors, Al-based junctions appear to be the most promising choice, owing to their small gap 
($\approx 45$ GHz in the bulk limit) and the flexibility of tuning the equilibrium gap by varying the 
thickness~\cite{Marchegiani2022,Connolly2024,McEwen2024,Nho2025,Weerdenburg2023,Cherney1969,Court2007}. The asymmetry can be further 
tuned by raising the temperature, since the smaller gap decreases more rapidly with increasing temperature. As shown in Fig.~\ref{Fig1}, 
we require long junctions with leads of lengths larger than the corresponding superconducting coherence lengths, along with high 
transparencies. We note that a compromise can be achieved by increasing the equilibrium gap asymmetry, which in turn allows 
the use of smaller transparencies. According to our numerical calculations (Ref.~\cite{Lahiri2025}, which considered two- and 
three-dimensional models, as well as this work), which are limited to a maximum equilibrium gap asymmetry $\Delta_{0,L}/
\Delta_{0,R}\approx 0.05-0.1$ due to computational bottlenecks (the Higgs resonance condition requires $eV=\Delta_{0,L}$, and 
smaller values of $eV$ necessitate more Floquet modes), in order to achieve $I_{2\omega_J}\sim I_{\omega_J}$ we require transparencies 
$\gtrsim 0.4$. Nevertheless, even at lower transparencies or lower asymmetry, when $I_{2\omega_J}< I_{\omega_J}$, the presence 
of Higgs renormalization can still be inferred from the resonant amplification of its signatures on tuning the voltage bias across 
the Higgs resonance condition $\omega_J=\omega_H$.

\let\oldaddcontentsline\addcontentsline
\renewcommand{\addcontentsline}[3]{}

\section{Acknowledgements}

This work was supported by the W\"urzburg-Dresden Cluster of Excellence ct.qmat, EXC2147, project-id 390858490, and the DFG (SFB 1170).
The authors gratefully acknowledge the scientific support and HPC resources provided by the Erlangen National High Performance Computing Center (NHR@FAU) of the Friedrich-Alexander-Universit\"at Erlangen-N\"urnberg (FAU) under the NHR project b169cb. NHR funding is provided by federal and Bavarian state authorities. NHR@FAU hardware is partially funded by the German Research Foundation (DFG) – 440719683.

\appendix

\section{Phenomenological model for Higgs excitement in Josephson junctions}
\label{appA}

In this appendix, we show how the Higgs mode is excited in a JJ using a toy model for the superconducting condensates of two tunnel 
coupled superconducting leads. In this simple analysis, we only look at a linear response calculation at the leading order in the 
Josephson coupling between the condensates of the two SCs. We thus start with the Euclidean action describing the effective field 
theory of the condensates, represented by a complex scalar field $\Psi$ minimally-coupled~\cite{Altlandbook} to a scalar potential 
field $V$. An equivalent analysis may be performed in the Keldysh space for the non-linear response. The action is given by
\begin{align}
S=&\int_{\tau,\mathbf{r}}\sum_{j=L/R} \bigg[ |(\partial_\tau+i2eV_j(\tau))\Psi_j|^2 + c^2|\nabla \Psi_j|^2\nonumber\\
 &\hspace{1.375cm}+ \frac{a_j}{2}|\Psi_j|^2+ \frac{u}{4}|\Psi_j|^4 \bigg]\nonumber\\
 - &\int_\tau J\big( \bar{\Psi}_{L,x=0}\Psi_{R,x=0} + \bar{\Psi}_{L,x=0}\Psi_{R,x=0} \big).
\end{align}
Here $V_L=V(\tau)$ and $V_R=0$, without loss of generality, capture the potential difference, $a_j\sim (T-T_{C,j})/T_{C,j}$ ($T_{C,j}$ 
is the critical temperature of the $j^{\text{th}}$ lead) in the superconducting phase, and $J$ is the Josephson coupling. Note that in 
Euclidean space, the minimally-coupled scalar potential, transforming like the time-like component of the electromagnetic four-potential, 
gets an additional factor of $i$ as well~\cite{Altlandbook}.

We use the ansatz $\Psi_j=(\Delta_{0,j}+h_j)e^{-i\vartheta_j}$ to separate the mean gap amplitudes $\Delta_{0,j}=\sqrt{-|a_j|/u}\Theta(-a_j)$
 of the isolated condensates, the Higgs fields $h_j$, and the complex phase factor $e^{-i\vartheta_j}$. The relevant terms in the 
 resulting action is obtained as
\begin{align}
S&=\int_{\tau,\mathbf{r}}\sum_{j=L/R}\bigg[ (\partial_\tau h_j)^2 + c^2(\nabla h_j)^2 + \frac{|a_j|}{2}|h_j|^2 \nonumber\\
&\hspace{18.3mm} + \Delta_{0,j}^2\big[ (\partial_\tau\vartheta_j-2eV_j(\tau))^2 + c^2(\nabla\vartheta_j)^2 \big]\bigg]\nonumber\\
&-\int_\tau 2J\Delta_{0,L}h_{R,x=0}\cos(\vartheta_{L,x=0}(\tau)-\vartheta_{R,x=0}(\tau))\nonumber\\
&-\int_\tau 2Jh_{L,x=0}\Delta_{0,R}\cos(\vartheta_{L,x=0}(\tau)-\vartheta_{R,x=0}(\tau)).
\end{align}
Here $|a_j|/2$ is the mass of $h_j$. The phase field fluctuations are massive, and thus the phase settles into its mean-field value, 
satisfying the Josephson relation $\partial_\tau\vartheta_j=2eV_j(\tau)$. Defining $X(\tau)=2J\Delta_{0,R}\cos(\vartheta_{L,x=0} 
-\vartheta_{R,x=0})$,
\begin{align}
\langle h_{L,i\omega_n,\mathbf{k}}\rangle=&\frac{X_{i\omega_n}}{\omega_n^2+c^2\mathbf{k}^2+\frac{|a_L|}{2}}.
\end{align}
On analytically continuing to real time
\begin{align}
\langle h_{L,\omega,\mathbf{k}}\rangle=&\frac{X_{\omega}}{-(\omega+i0)^2+c^2\mathbf{k}^2+\frac{|a_L|}{2}},
\end{align}
where $X_\omega$ is the Fourier transform of $\cos(\vartheta_{L,x=0}(t)-\vartheta_{R,x=0}(t))$. Note that along with $\tau\to it$, 
we also require $V\to -iV$~\cite{BenJacob1983,Altlandbook}. 

For a DC voltage bias, the Josephson phase $\phi=\vartheta_L-\vartheta_R$ satisfies $\partial_\tau\phi=2e(V_L-V_R)=2eV$. Now, we 
assume without loss of generality that $a_L\ll a_R$ (corresponding to $\Delta_{0,L}\ll \Delta_{0,R}$, which suppresses the Higgs 
field in the $R$ condensate). Hence,
\begin{align}
S&=\int_{\tau,\mathbf{r}}\sum_{j=L/R}\bigg[ (\partial_\tau h_j)^2 + c^2(\nabla h_j)^2 + \frac{|a_j|}{2}h_j^2 \bigg]\nonumber\\
&-\int_\tau 2J(\Delta_{0,L}h_{R,x=0}+h_{L,x=0}\Delta_{0,R})\cos(2eV\tau).
\end{align}
Clearly, we see that the Higgs field responds at frequency $2eV$, and the response is peaked at the Higgs mass $2eV=|a_L|/2$. 
Thus, we see that the potential difference $V$ provides a dynamic excitation to the Higgs field at frequency $2eV$, essentially 
arising from the interference of the two OPs whose phases differ by $2eVt$. This is embodied in the tunneling action 
$S_T=-\int_\tau 2Jh_{L,x=0}\Delta_{0,R}\cos(2eV\tau)$, which in Minkowski space becomes $S_T=-i\int_t 2Jh_{L,x=0}\Delta_{0,R}
\cos(2eVt)$. Note that the analysis of the regime $2eV>|a_L|/2$ requires an account of the damping of the Higgs mode, which we 
have neglected in this toy model.

\section{Pseudo-Floquet decomposition}
\label{appB}

Here we discuss the representation used in Eq.~\eqref{Grep}, which is not the usual Floquet decomposition. This is because, 
in general, there is no periodicity when $\omega_J/2$ and $\omega_r$ are incommensurate. We do, however, recover the periodicity 
for $\omega_r=0$ (or $\omega_J=0$). In this case, we can split the frequency integral into intervals of length equalling the Floquet frequency 
$\Omega=\omega_J/2$ as the periodic perturbation always scatters between two frequencies separated by $\Omega$ and never within 
a single interval. A two-point function which is periodic in the average argument, $G(t+T,t'+T)=G(t,t')\iff 
G(t_{\text{av}}+T,\delta t)=G(t_{\text{av}},\delta t)$ where $t_{\text{av}}=(t+t')/2$, admits the usual Floquet expansion
\begin{align}
G(t,t')=\sum_{m,n}\int_0^{\Omega} \frac{d\omega}{2\pi} &\overbrace{G(\omega+m\Omega,\omega+n\Omega)}^{G_{mn}(\omega)}\nonumber\\
&\hspace{0.7mm}e^{-i(\omega+m\Omega)t+i(\omega+n\Omega) t'},\label{usualfloq}
\end{align}
where $G_{mn}(\omega)$ satisfies
\begin{align}
&G\big((\omega+\Omega)+m\Omega,(\omega+\Omega)+n\Omega\big)=G_{(m+1)(n+1)}(\omega).
\end{align}
In this case, the pseudo-Floquet one given by Eq.~\eqref{Grep} and the usual Floquet one in Eq.~\eqref{usualfloq} are equivalent
\begin{align}
G(t,t')=&\sum_m\int_{-\infty}^\infty \frac{d\omega}{2\pi}G(\omega+m\Omega,\omega)e^{-i(\omega+m\Omega)t+i\omega t'}\label{pseudofloquetrep}\\
=&\sum_{m,k} \int_{0}^{\Omega}\frac{d\omega}{2\pi}G_{m0}(\omega+k\Omega)e^{-i(\omega+k\Omega+m\Omega)t+i (\omega+k\Omega) t'}\nonumber\\
=& \sum_{m,k} \int_{0}^{\Omega}\frac{d\omega}{2\pi}G_{(m+k)(k)}(\omega)e^{-i(\omega+(m+k)\Omega)t+i(\omega+k\Omega)t'}\nonumber\\
\xrightarrow[k\to n]{m+k\to m}&\sum_{m,n} \int_{0}^{\Omega}\frac{d\omega}{2\pi}G_{mn}(\omega)e^{-i(\omega+m\Omega)t+i(\omega+n\Omega)t'}\label{floquetrep}
\end{align}


\begin{thebibliography}{70}
\bibitem{sizejust} We have fewer sites in the right lead as $\delta\Delta_R(t)\approx 0$. Additionally, it has a larger equilibrium gap $\Delta_{0,R}>\Delta_{0,L}$, and hence, a shorter coherence length compared to the left lead ($\xi_{\text{sc},R}<\xi_{\text{sc},L}$).
\bibitem{nocosI} We do not find any significant $\cos(\phi(t))$ and $\cos(2\phi(t))$ currents.
\end{thebibliography}
\end{document}